\documentclass[a4paper,11pt]{article}
\pdfoutput=1
\usepackage{graphicx}
\usepackage{subfigure,amssymb,amsmath}
\usepackage{cite,color,url,hyperref}

\hypersetup{
   bookmarks=true,         
   unicode=false,          
   pdftoolbar=true,        
   pdfmenubar=true,        
   pdffitwindow=false,     
   pdfstartview={FitH},    
   pdfauthor={Author},     
   pdfsubject={Subject},   
   pdfcreator={Creator},   
   pdfproducer={Producer}, 
   pdfkeywords={keyword1} {key2} {key3}, 
   pdfnewwindow=true,      
   colorlinks=true,       
   linkcolor=blue,        
   citecolor=red,         
   filecolor=magenta,      
   urlcolor=cyan,           
   linktocpage = true,
   }

\newcommand{\comment}[1]{}
\newcommand{\newc}{\newcommand}

\def\issue(#1,#2,#3){{\bf #1}, #2 (#3)}

\def\PRL#1#2#3{{Phys. Rev. Lett.  \bf #1}, #3 (#2)}

\def\PREP(#1,#2,#3){Phys.\ Rep. \issue(#1,#2,#3)}
\def\EPJC(#1,#2,#3){Eur.\ Phys.\ J.\ C \issue(#1,#2,#3)}

\def\PLB{{\em Phys. Lett.}  B}

\def\r2{\sqrt 2}
\def\beq{\begin{equation}}
\def\eeq{\end{equation}}
\def\bea{\begin{eqnarray}}
\def\eea{\end{eqnarray}}
\def\sinW2{\sin^2\theta_W}

\def\mz2{M_{z}^2}
\def\c2b{\cos 2\beta}

\def\m#1{{\tilde m}_#1}

\def\mz{M_Z}
\def\m0{m_0}

\def\sec2w{sec^2\theta_W}

\def\tanbeta{{\rm tan}\beta}
\def\gmin2{(g-2)_\mu}

\catcode`\@=11 

\def\lsim{\mathrel{\mathpalette\@versim<}}
\def\gsim{\mathrel{\mathpalette\@versim>}}
\def\@versim#1#2{\vcenter{\offinterlineskip
    \ialign{$\m@th#1\hfil##\hfil$\crcr#2\crcr\sim\crcr } }}

\newc{\wt}{\widetilde}
\newc{\ra}{\rightarrow}
\newc{\s}{\smallskip}
\newc{\nn}{\noindent}
\newc{\non}{\nonumber}

\def \chonep{{\wt\chi_1}^{+}}
\def \chonem{{\wt\chi_1^-}}
\def \chonep2{{\wt\chi_2^+}}
\def \chonem2{{\wt\chi_2^-}}

\def \chonepm{{\wt\chi_1}^{\pm}}

\def \mchonepm{m_{\chonepm}}

\def \lspone{\wt\chi_1^0}
\def \mlspone{m_{\lspone}}
\def \lsptwo{\wt\chi_2^0}
\def \mlsptwo{m_{\lsptwo}}

\def\PRL{Phys. Rev. Lett.}

\def \lsptwo{\wt\chi_2^0}
\def \lspone{\wt\chi_1^0}
\def \chonem {{\wt\chi_1^\pm}}
\def \chargino2 {{\wt\chi_2^\pm}}

\def \ch2m {{\wt\chi_2^-}}
\def \lspone{\wt\chi_1^0}

\def \chonep {{\wt\chi_1^+}}

\def \PMET{p{\!\!\!/}_T}
\def \EMET{E{\!\!\!/}_T}
\def \PMETmin{p{\!\!\!/}_{T}^{\rm min}}
\def \EMETmin{E{\!\!\!/}_{T}^{\rm min}}

\DeclareMathAlphabet{\mathpzc}{OT1}{pzc}{m}{it}

\newcommand{\mst}[2][]{\ensuremath{m_{\tilde{t}_{{#2}}}^{{#1}}}}

\newcommand{\bsg}{{{\rm BR}(b\to s\gamma)}}
\newcommand{\bsmumu}{{{\rm BR}(B_{s}\to \mu^{+}\mu^{-})}}

\begin{document}


\begin{flushright}
{TIFR/TH/15-34}
\end{flushright}


\begin{center}

{\large \bf Status of the 98 - 125 GeV Higgs scenario with updated LHC-8 data} \\

\vskip 0.6cm
Biplob Bhattacherjee$^{a}$\footnote{biplob@cts.iisc.ernet.in}, 
Manimala Chakraborti$^{b}$\footnote{tpmc@iacs.res.in}, 
Amit Chakraborty$^{c}$\footnote{amit@theory.tifr.res.in}, \\ 
Utpal Chattopadhyay$^{b}$\footnote{tpuc@iacs.res.in},
and
Dilip Kumar Ghosh$^{b}$\footnote{tpdkg@iacs.res.in}
\vskip 0.6cm

{$^a$  Centre for High Energy Physics, \\ 
Indian Institute of Science, Bangalore 560012, India}

\vskip 0.1cm
{$^b$ Department of Theoretical Physics, \\
Indian Association for the Cultivation of Science, \\ 
2A \& B, Raja S.C.\,Mullick Road, 
Jadavpur, Kolkata 700\,032, India.} \\

\vskip 0.1cm
{$^c$  Department of Theoretical Physics,\\  
Tata Institute of Fundamental Research,\\
1, Homi Bhabha Road, Mumbai 400005, India}

\end{center}


\vskip 0.3cm

\begin{abstract}
In the context of minimal supersymmetric standard model (MSSM), 
we discuss the possibility of the lightest 
Higgs boson with mass $M_h = 98 $ GeV to be consistent with the 
$2.3\sigma$  excess observed at the LEP in the decay mode $e^+e^- \to Zh$, 
with $h \to b {\bar b}$. In the same region of the MSSM parameter space, 
the heavier Higgs boson $(H)$ with 
mass $M_H \sim 125 $ GeV is required to be 
consistent with the latest data on Higgs
coupling measurements at the end of 7 + 8 TeV LHC run with $25{\rm fb}^{-1}$
of data. While scanning the MSSM parameter space, we impose constraints 
coming from flavour physics, relic density of the cold dark matter as 
well as direct dark matter searches. We study the possibility of observing
this light Higgs boson in vector boson fusion process and associated 
production with $W/Z$ boson at the high luminosity $(3000~{\rm fb}^{-1})$ 
run of the 14 TeV LHC. Our analysis shows that this scenario
can hardly be ruled out even at the high luminosity run of the LHC. However,
the precise measurement of the Higgs signal strength ratios can play a major
role to distinguish this scenario from the canonical MSSM one. 

\end{abstract}


\newpage
\setcounter{footnote}{0}
\hrule
\tableofcontents
\vskip 1.0cm
\hrule

\vskip 1.0cm


\section{Introduction}
\label{intro}
The LHC has recently started its second phase of run. 
The discovery of {a new scalar particle }of mass
$\sim$ 125 GeV has been confirmed \cite{higgs}. The properties of this newly
discovered resonance seem to be in close agreement with that of the
Standard Model (SM) Higgs boson of the same mass \cite{Flechl}. 
The production modes for the Higgs boson at the LHC are those
via the gluon-gluon fusion, the vector boson ($W^{\pm}, Z$) fusion and the 
associated production with a vector boson ($W^{\pm}, Z$) or top quarks
whereas its most effective decay modes are into 4 leptons 
($e$ and $\mu$) 
, $W^+ W^{*-}$, and $\gamma \gamma $ channels. 
One of the main goals of the current 
LHC run with increased centre of mass energy and higher 
luminosity is to find out whether the discovered particle is
the Standard Model Higgs boson or
a part of an extended Higgs sector containing several other 
physical scalars. 

The Higgs sector of the Minimal Supersymmetric 
Standard Model (MSSM) \cite{SUSYreviews1,SUSYreviews2, 
SUSYbooks,djouadimssmrev} has a far richer spectrum than that of 
the SM. The model has two CP-even neutral
scalars (the lighter and the heavier ones $h$ and $H$ respectively), 
one CP-odd neutral scalar ($A$) and two charged 
scalars ($H^{\pm}$). 
At the tree level, only two input parameters other than the 
Z-boson mass ($M_Z$)
are required to specify the Higgs sector of the MSSM. These inputs are 
$(i)$ the mass of the pseudoscalar Higgs 
boson ($M_A$) and $(ii)$ the ratio of vaccum 
expectation values (vevs) of the two Higgs doublets of the MSSM ($\tan\beta$).
Dependence on other input parameters are induced once the 
radiative corrections { to the Higgs mass } are taken into account. 
We note that the couplings of $H$ to the gauge bosons
such as $H Z Z$, $H W^{+} W^{-}$ are proportional to 
$\cos (\beta -\alpha)$ where $\alpha$ is the Higgs mixing 
angle\cite{djouadimssmrev}. On the other hand, couplings 
like $h Z Z$, $h W^{+} W^{-}$ are proportional to $\sin (\beta -\alpha)$. 

In the well-known decoupling limit of the MSSM\cite{djouadimssmrev} characterized by the 
mass hierarchy $M_H \simeq M_A \simeq M_{H^{\pm}} \gg M_Z, M_h$, 
all the Higgs bosons become much heavier than the lightest one ($h$) 
making the latter 
to have SM-like mass as well as couplings.   
In this limit one has $\cos (\beta -\alpha) \rightarrow 0$ indicating 
negligibly small values for the aforesaid couplings of the heavier Higgs 
bosons.
Thus, it seems natural to consider the Higgs particle observed at the LHC
as the lightest CP-even Higgs of the MSSM in the decoupling limit.
However, the other possibility of having a non-decoupling regime of the MSSM
where the observed boson at 125 GeV is interpreted as the heavier CP-even 
Higgs scalar $H$ can be consistent with the non-decoupling regime 
of the MSSM Higgs bosons where $M_h \sim M_H \sim M_A \sim M_Z$\cite{djouadimssmrev}. 
In this case, all the MSSM
Higgses would be light with the lightest one lying somewhat below 125 GeV.
The above non-decoupling scenario may get its motivation from an old result 
by the LEP collaboration which corresponds to an excess of Higgs-like 
events around a mass of 98 GeV \cite{LEP98,Schael:2006cr}.
The excess was found in the channel 
$e^+ e^- \rightarrow Z h$ with $h$ decaying into $b \bar b$.
In a combined analysis of the four LEP working groups this excess reached 
a significance of 2.3$\sigma$. The above phenomena can 
not be explained within the SM since the SM Higgs boson would 
give rise
to a larger production cross-section. In our previous analysis of 
Ref.\cite{older} we explored the possibility of interpreting
the 2.3$\sigma$ excess events in the MSSM
with the name `Inclusive LEP-LHC Higgs (ILLH) 
scenario' where $h$ and $H$ are required
to correspond to the LEP excess near 98 GeV and the observed resonance 
at $\sim$ 125 GeV at the LHC respectively. 
There have been several studies in this direction and a partial list may be seen in Ref.\cite{
Drees:2005jg,Drees:2012fb,Christensen:2012ei,Christensen:2012si,Bhattacherjee:2012bu,Barbieri:2013nka,Cerdeno:2013qta,
Belanger:2012tt,Gunion:2012gc,Cerdeno:2013cz,Bomark:2014gya,Bomark:2015fga,
Badziak:2013bda,Allanach:2015cia,Astapov:2014mea}.
In spite of the fact that the above MSSM scenario is 
believed to be cornered in recent 
times \cite{Arbey:2014msa}, we believe that 
it is important to review the current status of the scenario 
in relation to the latest LHC data via focussing our attention 
in relevant zones of appropriate parameters of MSSM in a 
model independent way.  

We must note that if this scenario is indeed realized then
$(i)$ the value of $\sin^2 (\beta - \alpha)$ must be very small in order  
to explain the small $Z Z h$ coupling at LEP and $(ii)$ the couplings of $H$ must
be similar to that of the SM Higgs in order to be compatible with the 125 GeV
resonance observed at the LHC. The parameter region which satisfies these two
requirements must also pass through other direct constraints coming from 
the LHC, the most important ones are given as below. 
\begin{itemize}
\item
Exclusion limits in the channel $H/A \rightarrow \tau^+ \tau^-$ : Both 
the ATLAS and CMS collaborations have searched for a neutral 
Higgs-like boson $\Phi$ in the decay channel $\tau^+ \tau^-$ 
for certain benchmark scenarios \cite{Aad:2014vgg,CMS:2015mca}. 
However, in order to perform a model-independent 
analysis one must consider the bounds on $\sigma \times {\rm BR}(\Phi \to \tau^{+}\tau^{-})$ 
as a function of $m_{\Phi}$, where $\sigma$ denotes the production cross-section
for the non-minimal Higgs boson $\Phi$ 
decaying into the di-tau channel. Two different production modes for $\Phi$
are considered namely, the gluon-gluon fusion and the associated production with 
b-quark.

\item
Searches for $H^{\pm}$ : The ATLAS and CMS searches for $H^{\pm}$ are
performed in $t \bar t$ events with subsequent decays 
$t \rightarrow b H^{\pm}$ and $H^{\pm} \rightarrow \tau \nu$ 
\cite{TheATLAScollaboration:2013wia,CMS:2014cdp}. 
Model-independent upper bounds are obtained for 
${\rm BR}(t \rightarrow b H^{\pm}) \times {\rm BR}(H^{\pm} \rightarrow \tau \nu)$.
These searches exclude $\tan\beta$ up to 6 for 
90 GeV $< M_{H^\pm} <$ 150 GeV.
%
\end{itemize}
These two searches together restrict $\tan\beta$ to have a very narrow
range. Moreover, the constraints from heavy flavor physics also become
very important in this region of parameter space. In particular, 
experimental limits on
$\bsg$ and $\bsmumu$ are able to play very crucial roles to constrain the
MSSM parameter space under question.

In Ref.\cite{older} we imposed all of the above bounds
on the MSSM parameter space to probe whether it could accommodate 
the ILLH scenario. However, for the Higgs signal strength constraints
we used a conservative lower bound on ${\mu}^{\gamma \gamma}_{gg}$ 
(see Eq.\ref{rdef1}) taken as ${\mu}^{\gamma \gamma}_{gg} >0.5$.
Afterwards, the bounds from the 
ATLAS and CMS collaborations for the pseudoscalar and charged Higgs search 
channels have become more stringent. 
On the other hand, the Higgs signal strength results have become
more precise in recent times. Thus it 
seems very reasonable to reanalyze the 98-125 GeV Higgs scenario
in the light of the updated collider constraints and to check whether 
there is any room left in the MSSM framework to accommodate a 98 GeV
Higgs boson.

In this work we perform a detailed scan over the MSSM parameter space 
to find out the region allowed by all the relevant collider
constraints mentioned above. 
We also demand that the { MSSM parameter space must satisfy the } 
PLANCK limit on dark matter relic density.
Keeping these issues in mind we explore the possibility of
observing the signals of the ILLH scenario in the
high-luminosity run of the LHC. The plan of this paper is
as follows.

In Sec.~\ref{sec2} we discuss the major constraints
imposed on the MSSM parameter space of our interest and the parameter
ranges we choose to perform the scanning procedure. The impact
of the constraints on the MSSM parameter space as well as the main
features of the allowed zone are studied in Sec.~\ref{sec3}. In Sec.~\ref{sec4}
we analyze the prospects of the ILLH scenario in the high luminosity
run of the LHC. Finally, we conclude in Sec.~\ref{sec5}.



\section{Relevant constraints and parameter space scanning}\label{sec2}
In this section we enumerate the essential constraints and the scanning 
details of the MSSM parameter space considered in this analysis.
\subsection{The basic constraints for the ILLH scenario}
The ILLH scenario requires 
that the lightest CP-even Higgs boson should have a mass around 98 GeV, while 
the one observed at the LHC at 125 GeV to be the heavier 
CP-even Higgs boson. Thus, we 
consider the following ranges for $M_h$ and $M_H$.
\begin{eqnarray}
95~{\rm GeV} < M_h < 101~{\rm GeV}, \nonumber  \\
122~{\rm GeV} < M_H < 128~{\rm GeV}.
\label{lep-lhc-masslimits}
\end{eqnarray}
An uncertainty of 3 GeV in the Higgs boson mass is assumed which may come 
from the top mass uncertainty, uncertainties in the renormalization scheme 
and higher order loop corrections~\cite{higgsuncertainty3GeV}. 
The value of $\sin^2(\beta-\alpha)$ must lie within the following 
range in order to satisfy the LEP limit.
\begin{equation}
0.1<\sin^2(\beta-\alpha)<0.25.
\label{LEPcriterion}
\end{equation}

\noindent
\subsection{Constraints from the LHC}
The different production mechanisms of the Standard Model 
Higgs boson at the LHC are the
gluon-gluon fusion (ggF), vector boson fusion (VBF) and 
associated production with gauge bosons ($VH$, $V = W^{\pm}, Z$) 
or with a pair of top quarks ($t\bar t H$). Among its various possible 
decay modes, the decay into a pair of 
bottom quarks is the most dominant one. Other sub-dominant decay modes 
include final states involving a 
pair of SM gauge bosons ($VV^{*}$), 
$\tau^{+}\tau^{-}$ and 
$\gamma\gamma$ etc. The di-photon final states refer to loop-induced 
phenomena involving $W$-boson and heavy fermion loops.
Both the ATLAS and CMS collaborations have 
analyzed various production and decay modes of the Higgs boson observed 
at 125 GeV and put bounds on the various couplings of the 
SM Higgs. The signal strength parameter $\mu$ is defined 
as the ratio between 
the measured Higgs boson rate and its SM expectation as follows: 
\begin{equation}
{\mu}^{f}_{i} = \frac{\sigma_{i} \times {\rm BR}^{f}}
{{(\sigma_{i})}_{\rm SM} \times {({\rm BR}^{f})}_{\rm SM}}.
\label{rdef1}
\end{equation}
Here, $\sigma_{i}$ represents the production cross-section for 
a given new physics model with $i$= ggF, VBF, VH and $t\bar t H$ 
processes for a generic Higgs boson $H$ with 
$f = \gamma\gamma$, $ZZ^*$, $WW^{*}$, $ b\bar b$, $\tau^+ \tau^-$ being 
the decay modes of the Higgs boson. The subscript ``SM" 
represents the respective SM expectations. 
In Table \ref{tab:higgs}, we display the most updated 
combined results on various Higgs signal strengths by the ATLAS and 
CMS collaborations \cite{ATLAS-CMS-comb}. The subscript `F' denotes 
the combined data for the ggF and $t\bar t H$ process, while 
`V' signifies the combined VBF and VH processes. 
Even though the Higgs production through the ``fusion" (F) 
mode includes both the ggF and $t\bar t H$ processes, here 
we consider the ggF process only since 
$\frac{\sigma_{t\bar t H}}{\sigma_{ggF}} \sim$ 2\% as estimated 
by the combined ATLAS \& CMS data and also uncertainties in Higgs 
signal strength measurements associated to the $t\bar t H$ process 
being relatively large. 
\begin{table}[!ht]
\begin{center}
\begin{tabular}{|c|c|}
\hline
Channel  & Combined ATLAS + CMS signal strength \\
\hline
${\mu}^{\gamma\gamma}_{F}$ & $1.19^{+0.28}_{-0.25}$  \\
${\mu}^{\rm WW}_{F}$  & $1.0^{+0.23}_{-0.20}$   \\
${\mu}^{\rm ZZ}_{F}$  & $1.44^{+0.38}_{-0.34}$  \\
${\mu}^{\rm bb}_{F}$  & $1.09^{+0.93}_{-0.89}$  \\
${\mu}^{\tau\tau}_{F}$  & $1.10^{+0.61}_{-0.58}$ \\
\hline
${\mu}^{\gamma\gamma}_{V}$ & $1.05^{+0.44}_{-0.41}$  \\
${\mu}^{\rm WW}_{V}$  & $1.38^{+0.41}_{-0.37}$   \\
${\mu}^{\rm ZZ}_{V}$  & $0.48^{+1.37}_{-0.91}$  \\
${\mu}^{\rm bb}_{V}$  & $0.65^{+0.30}_{-0.29}$  \\
${\mu}^{\tau\tau}_{V}$  & $1.12^{+0.37}_{-0.35}$ \\
\hline
\end{tabular} 
\caption{Combined results of the Higgs coupling measurements by the ATLAS and 
CMS collaborations at the end of 7 + 8 {\rm TeV} run of the LHC with approximately
 25 ${\rm fb}^{-1}$ of data \cite{ATLAS-CMS-comb} }
\label{tab:higgs}
\end{center}
\end{table}
At the end of 8 TeV run of the LHC, the 
signal strength variables associated with the observed 
125 GeV Higgs boson still allow significant 
deviations from the SM predictions. Keeping this in mind, 
here we consider 2$\sigma$ deviations from the central 
value of various signal strength variables obtained after 
combining the ATLAS and CMS data.  
\begin{eqnarray}
 0.69 < {\mu}^{\gamma \gamma}_{F} < 1.75, \quad 0.6 < {\mu}^{WW}_{F} < 1.46, \quad 0.76 < {\mu}^{ZZ}_{F} < 2.2, \nonumber \\
-0.69 < {\mu}^{bb}_{F} < 2.95, \quad -0.06 < {\mu}^{\tau\tau}_{F} < 2.32, \quad 0.23 < {\mu}^{\gamma \gamma}_{V} < 1.93, \nonumber \\
 0.64 < {\mu}^{WW}_{V} < 2.2, \quad -1.34 < {\mu}^{ZZ}_{V} < 3.22, \quad 0.07 < {\mu}^{bb}_{V} < 1.25, \nonumber \\
0.42 < {\mu}^{\tau\tau}_{V} < 1.86. 
\label{tab:Rlhc}
\end{eqnarray}
Apart from the above, there are two experimental 
constraints that play crucial roles in the 
parameter space of our interest namely the limits from direct searches of 
the pseudo-scalar and charged Higgs bosons at the LHC. 
We note that from the 
direct searches of the pseudo-scalar Higgs boson
both the ATLAS and CMS collaborations have eliminated
the zone $90 <M_A < 250$~GeV 
for $\tan\beta \gsim$ 5.5 \cite{Aad:2014vgg,CMS:2015mca}. On the 
other hand, the ATLAS and CMS have also searched for light 
charged Higgs bosons using $t \bar t$ events via 
$t \rightarrow b H^+$ mode with 
$H^+ \rightarrow \tau^+ \nu_\tau$ 
\cite{TheATLAScollaboration:2013wia,CMS:2014cdp}. The ATLAS 
analysis indicates that the regions of 
parameter space with $\tan\beta$ between 2 to 6 
with $90 < m_{H^+} < 150$~GeV is disallowed. 
We note that the ATLAS and CMS exclusion limits are available for 
a few benchmark scenarios with specific 
choices of the MSSM model parameters (e.g. the so-called $m_h^{max}, 
m_h^{mod \pm}$ etc. scenarios\cite{Aad:2014vgg,CMS:2015mca,
TheATLAScollaboration:2013wia,CMS:2014cdp}). 
However, these scenarios seem to be rather over-simplified.
On the other hand, there exist model-independent limits on 
the production cross-section 
times branching ratio i.e., 
$\sigma \times {\rm BR}(\Phi \to \tau^{+}\tau^{-})$, for a 
non-standard Higgs boson $\Phi$ when it is produced via gluon-gluon 
fusion and b-quark associated processes \cite{Aad:2014vgg,CMS:2015mca}. We use these model-independent
limits in the present work. 

\subsection{Constraints from flavor physics and cosmological abundance of dark matter }
The light pseudo-scalar and charged Higgs bosons naturally make the
flavor physics constraints very significant for the parameter 
space of our interest. We consider the two most stringent 
rare b-decay constraints, namely $\bsg$ 
and $\bsmumu$, 
and allow $2\sigma$ deviation from the central limit\footnote{The current 
measurements of these two b-observables stand at 
${\rm BR}(B_s \to X_{s}\gamma) = 3.43 \pm 0.22 \pm 0.21({\rm theo.})$ 
and ${\rm BR}(B_s \to \mu^+ \mu^-) = 3.1 \pm 0.7 \pm 0.31 ({\rm theo.})$ \cite{Amhis:2014hma}. 
We follow Ref.~\cite{Kowalska:2015zja} for the conservative estimates of the theoretical 
uncertainties associated with these two flavor observables.} \cite{Amhis:2014hma}, 

\begin{equation}
2.82 \times 10^{-4}  <\bsg<4.04 \times 10^{-4},
\label{bsgammalimits}
\end{equation}
\begin{equation}
1.57 \times 10^{-9} <\bsmumu<4.63 \times 10^{-9}.
\label{bsmumulimit}
\end{equation}

Following the analysis of the PLANCK experiment\cite{Ade:2013zuv}
as in Ref.\cite{older}, we take the DM relic density limits as
$0.112 < \Omega h^2 <0.128$. However, in this
analysis we allow the possibility of having a multicomponent DM scenario. 
Hence, we consider only the upper limit of the relic density constraint
as given below.
\begin{equation}
\Omega_{{\widetilde \chi}_1^0}h^2<0.128.
\label{planckdata}
\end{equation}
This allows for the possibility of an under-abundant DM scenario
with relic density lying below the lower limit of the PLANCK data.
We also check the consistency with the upper bounds on the 
DM direct detection cross-section from the LUX \cite{Akerib:2013tjd} experiment. 

\subsection{Exploring the relevant MSSM parameter space for the ILLH scenario}
\noindent
Considering the data 
from $A/H \rightarrow \tau^+ \tau^-$ 
and $H^+ \rightarrow \tau^+ \nu_\tau$ search channels as constraints, we focus on the 
ILLH scenario for a small range of $\tan\beta$, namely 
$1<\tan\beta< 6.5$. 
Note that, LEP data \cite{Schael:2006cr} disfavors the region with 
$\tan\beta<3$ for SM-like SUSY Higgs boson search with SUSY breaking 
scale $M_{SUSY} = 1$ TeV. However, 
for a 98 GeV non-SM-like Higgs boson $h$, 
$\tan\beta$ can indeed be smaller than 3. Thus, in our 
analysis we probe the regions with smaller values of 
$\tan\beta$. On the other hand, choice of a model independent 
approach for Higgs mass motivates us for scanning up to $\tan\beta=6.5$, 
a value higher than the limit of the ATLAS and CMS data. 

We choose the decoupling zone ($\sim$ 3 TeV) 
for the first two generations of squarks and 
sleptons, considering the fact that there is no effect on the Higgs 
spectra in phenomenological MSSM (pMSSM) of the above scalars\cite{Djouadi:1998di}.


The parameter range over which we perform random scan 
can be listed as follows.
\begin{eqnarray}
1<\tan\beta <6.5, ~0.12< M_A < 0.3 ~{\rm TeV}, ~0.3~{\rm TeV} <\mu <
12~{\rm TeV}\footnotemark  \nonumber \\
0.05~{\rm TeV} < M_1,  M_2 < 1.5~{\rm TeV}, ~0.5~{\rm TeV} <M_3 <3~{\rm TeV}, \nonumber \\
-8 ~{\rm TeV} < A_t, A_b < 12 ~{\rm TeV},  ~A_u=A_d=A_{\tau}=A_e=0, \nonumber \\
0.3~{\rm TeV} <M_{\tilde q_3} <  5 ~{\rm TeV}, ~{\rm where}, ~{\tilde q_3}
\equiv {\tilde t_L},{\tilde t_R},{\tilde b_L},{\tilde b_R} \nonumber \\
M_{{\tilde q}_i}= 3~{\rm TeV}, {\rm for}~i=1,2~~ {\rm and}~~ M_{{\tilde l}_i}=3~{\rm TeV}, {\rm for}~i=1,2,3.
\label{parameterRanges}
\end{eqnarray}
\footnotetext{In our quest to explore the validity of the ILLH scenario we 
include large values $\mu$ in our parameter scan, keeping aside any 
fine-tuning related concern.}
The relevant SM parameters are chosen as, 
${m_b}^{\overline {\rm MS}}(m_b)=4.19$~GeV and $m_t^{\rm pole}=173.3 \pm 2.8$~GeV 
(a larger error amount is considered following the argument of 
Ref.\cite{Alekhin:2012py}) and the strong coupling constant 
$\alpha_{s}(M_{Z}) = 0.1172$. Stringency of satisfying  
Eq.\ref{lep-lhc-masslimits} to Eq.\ref{planckdata} or primarily 
Eq.\ref{lep-lhc-masslimits} and Eq.\ref{LEPcriterion} requires a very 
dense parameter scan. In this work, the number of parameter points scanned is 
more than 80 million over the above ranges.

The publicly available code SuSpect (version 2.43) \cite{suspect} is 
used for spectrum generation and micrOmegas (version 3.6.9.2) \cite{micromegas} 
is used for calculating the relic density and flavor observables
while the branching ratios of the 
Higgs bosons are computed via HDECAY \cite{hdecay}.
We calculate the 
Higgs production cross-section using 
SuShi \cite{sushi}. 
The lower limits on the sparticle 
masses are imposed from the LEP and LHC data. 
We consider the lightest top and bottom 
squark masses are greater than 500 GeV, while the 
gluinos are assumed to be heavier than 1.4 TeV \cite{PDG}. 
We also impose the LEP limit on the 
lightest chargino mass to be 100 GeV \cite{PDG}.   
The charge color breaking (CCB) constraints are already 
imposed by SuSpect while scanning the parameter 
space\footnote{Since the parameter ranges are associated with large values of
$\mu$ and $A_t$ we have further used a more dedicated check for the CCB 
constraints by using the code Vevacious (version 1.1.3)\cite{Camargo-Molina:2013qva} for
the two chosen benchmark points (BPs) (see 
Sec.~\ref{sec3}). The code is able 
to avoid a CCB minima or it can check cosmological stability in 
presence of a CCB vacuum via using the code 
CosmoTransitions\cite{Wainwright:2011kj}. See 
Ref.\cite{Chattopadhyay:2014gfa} and references therein for further details.}.


\section{Result}
\label{sec3}
\begin{figure}[!htb]
\includegraphics[angle =0, width=0.50\textwidth] {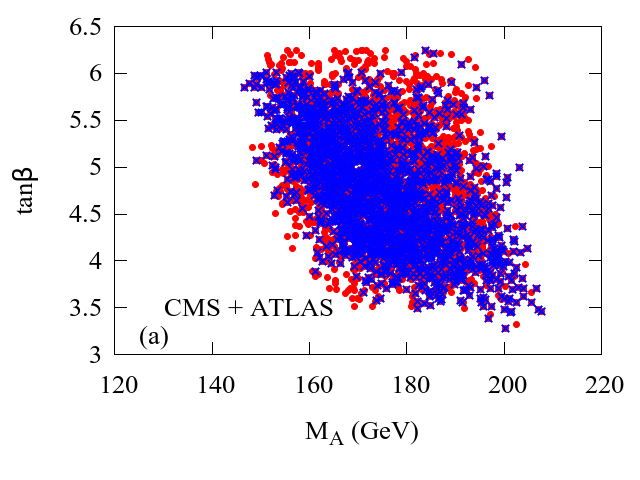}
\includegraphics[angle =0, width=0.50\textwidth] {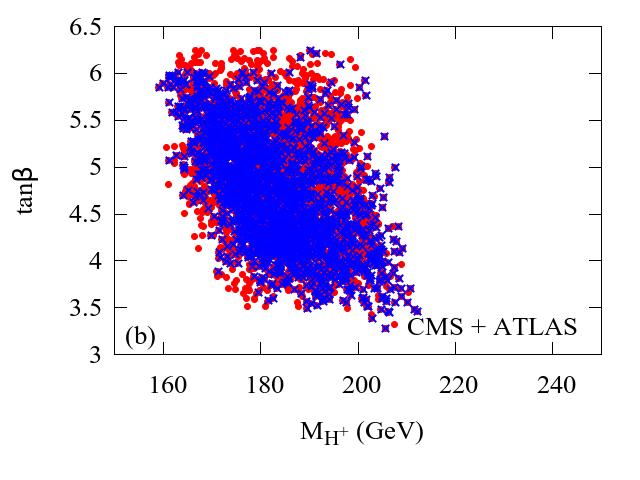} 
\caption{ {\it Scatter plot in the $M_{A} - \tan\beta$ (left) and 
$M_{H^{\pm}} - \tan\beta$ (right) plane. The red circles 
represent the points which satisfy all the constraints 
(Eqs.~\ref{lep-lhc-masslimits}-\ref{bsmumulimit} ), while 
blue crosses display points which in addition satisfies the DM relic density 
constraint of Eq.~\ref{planckdata}. }} 
\label{fig:ma_tanb}
\end{figure}

In Fig.~\ref{fig:ma_tanb}, we display the allowed parameter space in 
the $M_{A} - \tan\beta$ and $M_{H^{\pm}} - \tan\beta$ planes, where 
the red circles represent the points which satisfy all the constraints 
(Eqs.~\ref{lep-lhc-masslimits}-\ref{bsmumulimit}) except Eq.~\ref{planckdata},
while blue crossed points satisfy Eq.\ref{planckdata}. 

\begin{figure}[!htb]
\includegraphics[angle =0, width=0.50\textwidth] {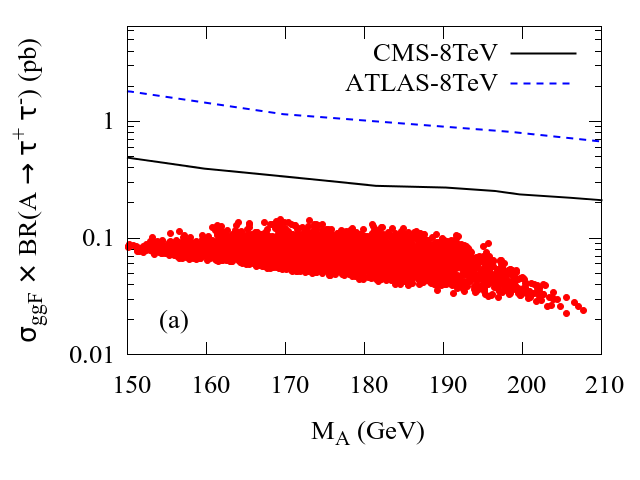}
\includegraphics[angle =0, width=0.50\textwidth] {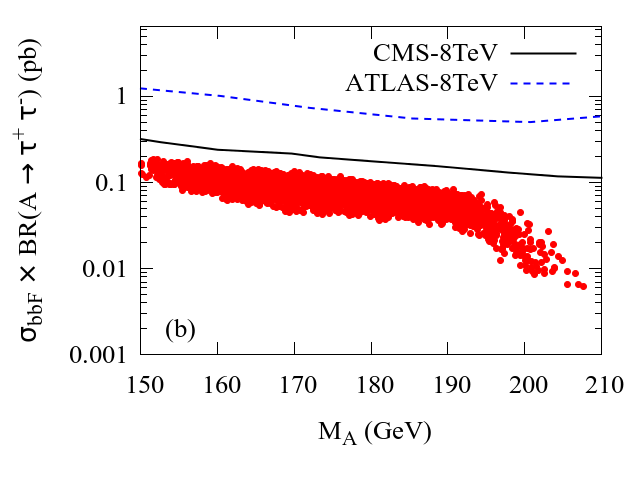}
\includegraphics[angle =0, width=0.50\textwidth] {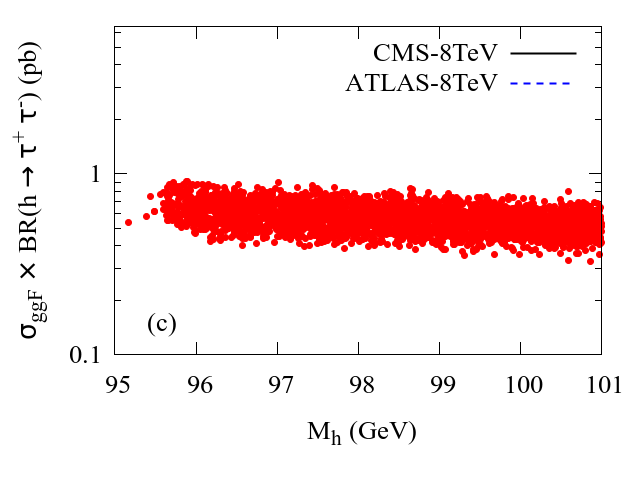} 
\includegraphics[angle =0, width=0.50\textwidth] {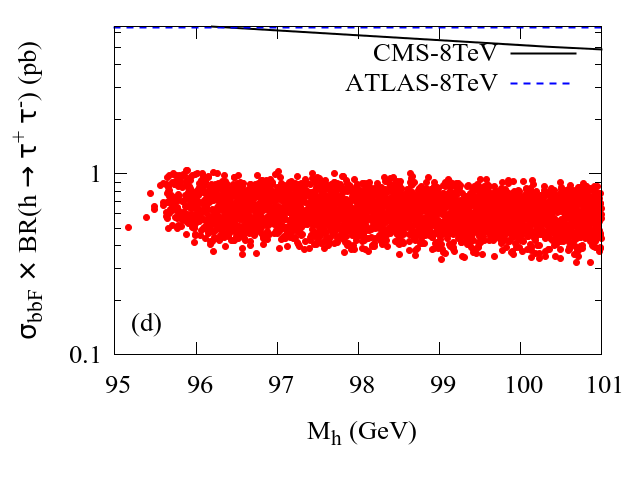} 
\caption{ {\it Distribution of $\sigma \times {\rm BR}(\Phi \to \tau^{+}\tau^{-})$
for all the red circled points corresponding to the Fig.~\ref{fig:ma_tanb},
i.e., for those points which satisfy all the constraints 
(Eq.\ref{lep-lhc-masslimits}-\ref{bsmumulimit}) except the relic density 
bound of Eq.\ref{planckdata}. The black solid and blue
dashed lines represent the CMS and ATLAS upper limits on this quantity. The 
upper and lower plots in the left panel assume Higgs is produced 
via ggF process, while bbF production mechanism is considered in the 
right panel. For details see text.}} 
\label{fig:mphi_sigmaBR}
\end{figure}

\vskip 0.3cm 
In Fig.~\ref{fig:mphi_sigmaBR}, we 
show the distribution of $\sigma \times {\rm BR}(\Phi \to \tau^{+}\tau^{-})$ 
where $\Phi \equiv h, A$, with respect to $M_h$ and $M_A$ for all the red circled points 
corresponding to Fig.~\ref{fig:ma_tanb}, 
i.e., for those points which satisfy all the 
constraints (Eqs.~\ref{lep-lhc-masslimits}-\ref{bsmumulimit}) except 
Eq.~\ref{planckdata}. 
The black solid and blue 
dashed lines represent the CMS and ATLAS bounds respectively. The ggF 
production cross-section for both $h$ and $A$ increases 
with decrease of $\tan\beta$. However, the branching ratio of 
$h,A \to \tau^{+}\tau^{-}$ 
also gets reduced as $\tan\beta$ decreases. Thus, for 
low values of $\tan\beta$ as in the region of our interest, the 
product $\sigma(ggF) \times {\rm BR}(\Phi \to \tau^{+}\tau^{-})$
is still below the present 
experimental sensitivity, as can be seen from the left figures
of both the upper and lower panels. However, when one considers 
the production mode of the Higgs boson in association 
with $b\bar b$, interestingly the present 
exclusion bounds are found to be very close to the model predictions. 
A better sensitivity in the $b\bar b$-fusion channel results in 
strong bounds on our parameter space. A sudden fall in 
$\sigma \times {\rm BR}(\Phi \to \tau^{+}\tau^{-})$ distribution 
near $M_{\phi} = 190~{\rm GeV}$ originates from the 
opening of the dominant decay mode $A \to Z{h}$ 
($m_{h} \sim 98~{\rm GeV}$ in our case) and consequent strong reduction 
in the branching ratio of $\Phi \to \tau^{+}\tau^{-}$. Thus, 
a closer look at these distributions reveals that
one or two orders of improvement in the measurement of the 
quantity $\sigma \times {\rm BR}(\Phi \to \tau^{+}\tau^{-})$ for both the production 
processes will put strong constraint on the ILLH 
parameter space. 

\vskip 0.5cm

The allowed points in the present scenario correspond to
the charged Higgs boson mass lying in the range 
160 - 200 GeV. Thus, the dominant decay modes 
are seen to be $H^{\pm} \to \tau \nu_{\tau}$ and/or $H^{\pm} \to t\bar{b}$. 
Both the ATLAS and CMS collaborations have performed 
searches for the charged Higgs bosons with masses larger than that of the top 
quark \cite{Aad:2014kga,CMS:2014pea}. We find that the 
low $\tan\beta$ region with $M_{H^{\pm}} \ge 175$~GeV is consistent 
with the current LHC data.

\begin{figure}[!htb]\centering 
\includegraphics[angle =0, width=0.5\textwidth] {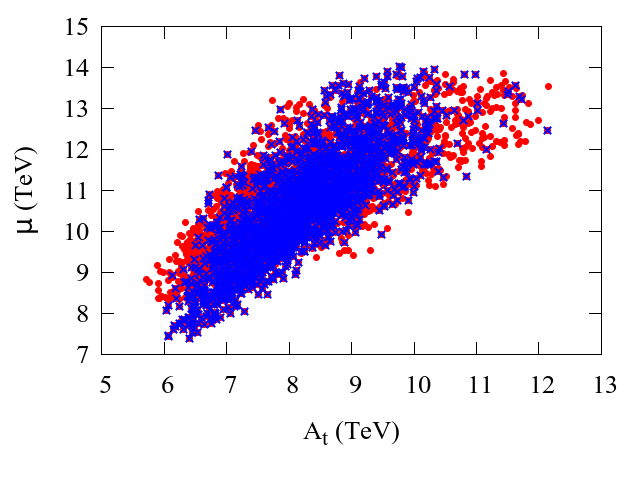}
\caption{ {\it Plot in $A_{t} - \mu$ plane. The symbols carry the same 
meaning as in Fig.~\ref{fig:ma_tanb}. }}
\label{fig:atmu}
\end{figure}


\subsection{Light charged Higgs bosons and flavour data}
Flavor observables play a crucial role in determining the 
valid regions of the MSSM parameter space. For example, 
rare B-decays that are helicity suppressed within the SM may on the other hand 
receive large contributions from the loop corrections involving 
SUSY particles. 
Two such rare B-decays are the radiative decay $\bsg$ 
and the 
pure leptonic decay $\bsmumu$. We will outline the relevant 
points of these constraints pertaining to our scenario with light $H^\pm$ 
while having large $\mu$ and large third generation of trilinear coupling
parameters.  We will particularly focus on the features of the valid 
parameter zones as allowed by the above constraints.  In our analysis with 
positive $\mu$ and gluino mass ($M_{\tilde g}$) it turns out that the valid 
zones also have positive $A_t$.  This may be seen in the $A_t-\mu$ 
plot of Fig.~\ref{fig:atmu}.  The red circles represent the points which
satisfy all the constraints of 
Eq.~\ref{lep-lhc-masslimits} to \ref{bsmumulimit}, 
while blue crosses 
indicate parameter points that additionally satisfy the DM relic density 
constraint of Eq.~\ref{planckdata}.   

The experimental data of $\bsg$ leaves a very small room for any Beyond the 
Standard Model (BSM) contribution. SUSY scenarios are effectively 
constrained by $\bsg$ (which has both an upper and a lower limit) 
due to cancellation of relevant diagrams, when the individual 
SUSY contributions may become large. However, we will see the importance of 
next-to-leading order contributions in regard to this constraint.  
In the SM, the $t-W$ loops cause non-zero contributions to $\bsg$, almost saturating the experimental value. 
In the MSSM, the dominant contributions to $\bsg$ come from the 
$t-H^\pm$ and ${\tilde t}_{1,2}-{\tilde \chi}_{1,2}^\pm$ 
loops\cite{bsgammaSMoriginals,bsgammaSUSYorigEtc}, 
where the former type of loops comes with the same sign with that of the $t-W$ loops of the SM. 
Considering the contributions of $\chonepm-\tilde t_1$ loops one has\cite{Carena:2000uj}:
\begin{equation}
\left.
\bsg \right|_{\chi^{\pm}} \propto
\mu A_t \tan\beta f(\mst{1},\mst{2},m_{\tilde{\chi}^+})
\frac{m_b}{v (1 + \Delta m_b)} .
\label{chloop}
\end{equation} 

On the other hand, for the $t-H^\pm$ loop contributions one finds\cite{Carena:2000uj}
\begin{equation}
\left.
\bsg \right|_{H^+}
\propto \frac{\left( h_t \cos\beta  - \delta h_t
\sin\beta \right)}{v \cos\beta} \;
g(m_{H^+},m_t) \frac{m_b}{(1 + \Delta m_b)}.
\label{hcloop}
\end{equation}
Here $f$ and $g$ of Eqs.~\ref{chloop},\ref{hcloop} are the loop functions. 
$\Delta m_b$ refers to the SUSY corrections to bottom mass where the SUSY QCD (SQCD)
corrections may have a significant role which we will discuss soon.    
$\delta h_t$ as appearing in the
second term of Eq.~\ref{hcloop} principally 
results from the corrections to the top
quark Yukawa coupling due to SQCD effects and this gives rise to a 
next-to-leading order (NLO) effect in $\bsg$. 
The dominant SQCD corrections to $\delta h_t$ arising from the 
gluino-squark loops are given by\cite{Carena:2000uj},   
\begin{equation}
\delta h_t  = h_t \frac{2 \alpha_s}{3 \pi}\mu M_{\tilde{g}}
\left(
\cos^2\!\theta_{\tilde{t}} I(m_{\tilde{s}_L},m_{\tilde{t}_2},
M_{\tilde{g}})
+
\sin^2\!\theta_{\tilde{t}} I(m_{\tilde{s}_L},m_{\tilde{t}_1},
M_{\tilde{g}})  \right) ,
\label{deltaht}
\end{equation}
where $I$ is again a loop function and $\theta_{\tilde{t}}$ is the squark 
mixing angle for the third generation. 

Coming to the SQCD corrections to $\Delta m_b$\footnote{SUSY electroweak corrections to 
the bottom Yukawa couplings can also be large for large 
$\mu$\cite{Carena:1999py}} 
we have two types of contributions namely, from the ${b- \tilde g}$ and 
$\chonepm-t$ loops\cite{Carena:1999py,Anandakrishnan:2014qxa,Carena:1994bv,Haisch:2012re,Degrassi:2000qf}. 
Following Ref.\cite{Anandakrishnan:2014qxa} 
the corrections are given as below.
\begin{equation}
\Delta m_b\simeq \frac{2 \alpha_s}{3 \pi} M_{\tilde{g}} (\mu\tan\beta-A_b)
I(m_{\tilde{b}_1},m_{\tilde{b}_2},
M_{\tilde{g}})
+\frac{h_t^2}{16\pi^2}\mu (A_t\tan\beta -\mu)
I(m_{\tilde{t}_1},m_{\tilde{t}_2},\mu).
\label{botsqcd}
\end{equation}


One finds that over the parameter space scanned, the 
next-to-leading order (NLO) effects arising from the SQCD 
corrections of 
the top Yukawa coupling (gluino-squark loop diagrams) 
that in turn affects the contributions from  
the $t-H^\pm$ loops may have significant role in $\bsg$.  
Typically, these NLO corrections are known to be important for 
large values of $\tanbeta$. But in spite of $\tan\beta$ being 
small in our analysis, the same corrections are also 
very important because of possible large values of 
$\mu$\cite{Carena:1999py} considered in this work.
The reason is that these NLO corrections are approximately proportional to 
$\mu M_{\tilde g}\tan\beta$\cite{Carena:2000uj,Carena:1999py}.  Thus,
regions of parameter space with large values of $\mu$ may change the 
$\bar t b H^{+}$ coupling leading to reduction of $t-H^{\pm}$ loop
contributions to $\bsg$ \cite{Carena:2000uj}. Furthermore, as seen 
in Eq.~\ref{botsqcd} the SQCD corrections to $\Delta m_b$ 
can be substantially 
large in spite of the fact that $\tan\beta$ is small in our analysis.
This will have an overall suppression effects of the aforesaid SUSY 
loop contributions to $\bsg$ over the valid parameter space of interest 
with large and positive values of $\mu$ and $A_t$.  
We note that the valid regions of parameter space that satisfy 
$\bsg$ correspond to heavy enough SUSY spectra and do not 
involve cancellation between the two basic types of 
SUSY loop diagrams. We have further imposed the 
constraint from $\bsmumu$
\cite{Aaij:2013aka, Chatrchyan:2013bka}
in this analysis. In the 
parameter space that survives after imposing the constraint from 
$\bsg$ it turns out that, 
$\bsmumu$ does not take away 
any significant amount of parameter space because of its smaller SUSY 
contributions. This arises out of cancellation of relevant terms 
for the positive sign of $\mu A_t$ that gives rise to a 
positive value for the dimensionless Wilson coefficient 
of the semileptonic pseudoscalar operator 
(See Ref.\cite{Haisch:2012re} and references therein). 



\subsection{Benchmark points}

\begin{table}[!htb]
\parbox{.45\linewidth}{
\centering
\small{
\begin{tabular}{|c|c|c|}
\hline
Point & BP1 & BP2  \\
\hline
\hline
\multicolumn{3}{|c|}{Input Parameters} \\
\hline
\hline
$\tan\beta$ &  4.28     &  5.22\\
\hline
$\mu$ (GeV)&    9333 &      9177 \\
\hline
$M_1$ (GeV) &    731.2 &    600.4 \\
\hline
$M_2$ (GeV)&   493.3 &   933.4 \\
\hline
$M_3$ (GeV)&  2056.7 &   3298 \\
\hline
$A_t$ (GeV)&  7047.2 &    6994.9\\
\hline
$A_b$ (GeV)&  -4838.3 &  -2560.4 \\
\hline
$M_{\tilde Q_{3L}}$(GeV) &  1473.7 & 1522.8 \\
\hline
$M_{\tilde t_{R}}$(GeV)  &  2806.9 &   3132.4 \\
\hline
$M_{\tilde b_{R}}$ (GeV) &  671.6 &  536.4 \\
\hline
$m_t$ (GeV)  &  173.1 &   174.3\\
\hline
\end{tabular}
}
\caption{
Input parameters for the two benchmark points 
allowed by the constraints 
Eqs.~\ref{lep-lhc-masslimits}-\ref{planckdata}.}
\label{bptable1}
}
\hfill
\parbox{.45\linewidth}{
\centering
\small{
\begin{tabular}{|c|c|c|}
\hline
Point & BP1 & BP2 \\
\hline
\hline
\multicolumn{3}{|c|}{Mass spectrum} \\
\hline
\hline
$M_h$ (GeV)  &  97.8&    96.4\\
\hline
$M_H$ (GeV)  &  126.5 &    127.9 \\
\hline
$M_{H^{\pm}}$ (GeV)  & 191.8 &  175.02    \\
\hline
$M_A$ (GeV) &  185.7 &  166.3 \\
\hline
$m_{\tilde g}$ (GeV)& 2230.6   &  3355.8\\
\hline
$m_{\tilde t_1}$ (GeV)& 689.5 & 913.7  \\
\hline
$m_{\tilde t_2}$ (GeV)&  1409.2 &   2232.4\\
\hline
$m_{\tilde b_1}$ (GeV)&  793.4& 614.5 \\
\hline
$m_{\tilde b_2}$ (GeV)&   1560.9 &  1596.5 \\
\hline
$\mchonepm$  (GeV)&   492.9 &   933.0 \\
\hline
$\mlspone$ (GeV)   &  492.9 & 600.4  \\
\hline
$\mlsptwo$ (GeV)  &  731.1 &   933.0 \\
\hline
\end{tabular}
}
\caption{
Mass spectrum for the two benchmark points.}
\label{bptable2}
}
\end{table}

\hspace*{1.5cm}

\begin{table}[!htb]
\begin{center}
\small{
\begin{tabular}{|c|c|c|}
\hline
Point & BP1 & BP2 \\
\hline
\hline
\multicolumn{3}{|c|}{Values of the Observables} \\
\hline
\hline
$\bsg \times 10^{4}$ & 3.65   & 3.85 \\
\hline
$\bsmumu \times 10^{9}$ & 2.67 &  2.21 \\
\hline
$\Omega_{\tilde \chi_1}h^2$  & 0.006  & 0.05\\
\hline
${\mu}^H_{gg}(\gamma \gamma)$   &   1.00 & 1.35 \\
\hline
$\sigma_{\tilde \chi p}^{\rm SI} \times 10^{10}$ (pb) & 1.27 & 0.07 \\
\hline
${\mu}^H_{gg}(ZZ)$   &  0.93 &     1.31 \\
\hline
${\mu}^H_{gg}(W^+W^-)$ &    0.89 &    1.25 \\
\hline
${\mu}^H_{Vh/H}(b \bar b)$ &   0.57 &    0.49\\
\hline
${\mu}^H_{Vh/H}(\tau^+ \tau^-)$  &   1.23 &  1.38  \\
\hline
\end{tabular}
}
\end{center}
\label{bptable}
\caption{ A few relevant observables for the two 
benchmark points considered in Table~\ref{bptable1}.}
\end{table}

We show Tables~2-4 for the choice of 
two benchmark points (BP1 and BP2) allowed by the constraints from 
Eq.~\ref{lep-lhc-masslimits} to \ref{planckdata}. Apart from the 
above constraints, large values of $\mu$ and $A_t$ in the valid region 
of parameter space specially motivate us  
in analyzing the effect of imposing the CCB 
constraints in a general setup going beyond the traditional constraints 
of CCB (see Ref.\cite{Chattopadhyay:2014gfa} and references therein 
for details) as used in the code 
SuSpect. We particularly analyze the above for only these two BPs rather 
than the entire parameter space simply for economy of computer time. 
We analyze the BPs by considering non-vanishing vacuum expectation values 
(vev) for the Higgs scalars and top-squarks. 
BP1 corresponds to a stable vacuum. 
BP2 has a cosmologically long-lived vacuum while allowing 
quantum tunneling.  Moreover, the BPs are so chosen
that the squark masses of the third generation lie above $\sim$ 600 GeV,
sufficiently large to be safely above the current bounds from the LHC. 
For BP1 $\mchonepm \simeq \mlspone$ and
$\lspone$, which is the lightest SUSY particle (LSP) in our case, 
is almost wino-like. Thus, the mechanisms which lead to right relic
density of the DM are mainly $\lspone-\chonepm$ coannihilation and
$\chonepm$ mediated $\lspone$ pair-annihilation to $W^{\pm}$. 
The strong annihilation and coannihilations make the LSP underabundant. 
$\mchonepm$ is taken to be
above 270 GeV so as to be consistent with the results from the
disappearing charged tracks searches at the LHC (which impose a
lower limit of 270 GeV for the masses 
of the wino-like LSPs\cite{disaptrack}).
The LSP in BP2 is 
predominantly a bino. The
main annihilation mechanism in this case is the
annihilation of sbottom pairs into a pair of gluons
in the final state. The spin-independent $\lspone$-proton scattering 
cross-section $\sigma_{\tilde \chi p}^{\rm SI}$ is also
seen to be well below the limit provided by the LUX experiment in
both the cases. The smallness of 
$\sigma_{\tilde \chi p}^{\rm SI}$ arise as a result of
large values of $\mu$ considered  in this study leading to 
a negligible higgsino component within the LSP \cite{Chakraborti:2014fha}.


\section{Prospects at the high luminosity run of the LHC}
\label{sec4}
In the last section we studied the available parameter 
space in the MSSM consistent with the ILLH scenario and also 
provided with two benchmark points allowed by the 
LHC as well as low energy physics data. In this section, we 
proceed to discuss the sensitivity of the high luminosity 
run of the LHC to probe the ILLH scenario. 
We start with the possibility to discover/exclude a 98 GeV Higgs boson 
produced through the vector boson fusion (VBF) and 
Higgsstrahlung processes. Note that, the reason behind our 
choice of the two above-mentioned processes is that the 
Higgs boson production cross-section is directly proportional 
to the Higgs to gauge boson coupling $\sin^{2}({\beta - \alpha})$. Moreover, 
to satisfy the LEP excess we require $\sin^{2}({\beta - \alpha}) \sim $0.2. 
Thus, if we observe a Higgs boson with mass $\sim$ 100 GeV 
in the associated/VBF processes with cross-sections 
$\sim$ 20\% of the SM cross-section, that can be thought of as a 
smoking gun signal of the ILLH scenario. Furthermore, we 
also analyze how the future precision measurements of various 
Higgs signal strength variables may be used as 
an indirect probe for the ILLH scenario.

\subsection{Direct search: 1. Vector boson fusion process}

The Vector Boson Fusion (VBF) process, $pp \to jj H$, (where $j$ stands
for light jets) is one of the 
most promising channels for the measurement of various properties 
of the observed 125 GeV Higgs boson at the LHC. It is a t-channel 
scattering process of two initial-state quarks with each one
radiating a $W/Z$ which further annihilates to produce a 
Higgs boson. Characteristic features of this 
process are the presence of two energetic 
jets with a large rapidity gap along with a large invariant mass 
and the absence of a significant amount of hadronic activity in the
central rapidity region. Even though the ggF process is the 
dominant production mechanism for the Higgs boson, 
due to the above distinctive features, VBF is  
sensitive enough to a precise measurement of various properties of 
the observed Higgs boson.

The ATLAS collaboration estimated the 
sensitivity of the VBF process for low mass 
Higgs bosons ($M_{H} < 130$~GeV), where $H \to \tau^+\tau^-$ decay
mode was considered at the 14 TeV LHC \cite{Aad:2009wy}. 
In the present work we perform a collider analysis following 
the ATLAS simulation to probe the discovery 
potential of the 98 GeV Higgs boson at the LHC-14. The 
ATLAS simulation considered three different 
decay modes of $\tau$, namley $\tau_{\ell}\tau_{\ell}$, 
$\tau_{\ell}\tau_{h}$, $\tau_{h}\tau_{h}$ where 
$\tau_{\ell}$ and $\tau_{h}$ denote the leptonically and 
hadronically decaying $\tau$-leptons respectively. From 
their analysis it is evident that the $\tau_{\ell}\tau_{h}$ channel 
has the best sensitivity compared to the other two possible 
modes (see Fig.~16 of Ref.~\cite{Aad:2009wy}). Thus, in this 
work we confine ourselves in the $\tau_{\ell}\tau_{h}$ channel only. 
Note that, even though we follow the ATLAS simulation for our analysis, 
we further vary the selection cuts to optimize the signal to 
background ratio.

In order to tag the $\tau$-leptons as $\tau$-jets, 
we first identify $\tau$ through its hadronic 
decays and then demand 
that the candidate jet must have $|\eta|<$ 2.5 and 
$p_{T} >$ 30 GeV. Besides, the jet must also 
contain one or three charged tracks with 
$|\eta_{\rm track}| < $ 2.5 with the highest track 
$p_{T} > $ 3 GeV. To ensure 
proper charged track isolation, we additionally demand 
that there are no other charged tracks with 
$p_{T} > $ 1 GeV within the candidate jet. 
The di-tau invariant mass is calculated using the ``collinear 
approximation technique" assuming the $\tau$-lepton 
and its decay products to be collinear \cite{Elagin:2010aw} 
and the neutrinos to be the only source\footnote{An alternative 
technique preferred by the experimental collaborations is 
the Missing Mass Calculator (MMC) method for the reconstruction of 
$\tau\tau$ invariant mass \cite{Elagin:2010aw}. In this paper, 
however, we restrict ourselves to the ``collinear 
approximation technique".} of $\EMET$. Neglecting the 
$\tau$ rest mass, the di-tau invariant mass can be written as, 
      \begin{equation}
        m^{2}_{\tau\tau} = 2 (E_{h} + E_{\nu h})(E_{\ell} + E_{\nu \ell})
        ( 1 - \cos{\theta_{\ell h}}), 
      \end{equation}
where $E_h$ and $E_\ell$ represent the total energy of the 
hadronically and leptonically decaying $\tau$s respectively, while 
$\theta_{\ell h}$ 
represents the azimuthal angle between the directions associated with
the above two decay modes of the $\tau$-lepton. We can now 
introduce two dimensionless variables $x_{\ell}$ 
and $x_h$, the fraction of $\tau$'s momentum taken away by the 
visible decay products, and rewrite the di-tau invariant mass 
as follows (with $x_{\ell,h} >0$).
\beq
m_{\tau\tau} = \frac{m_{\ell h}}{\sqrt{{x_{\ell}}{x_{h}}}},
\eeq
where
$m_{\ell h}$ is the invariant mass of the visible $\tau$-decay 
products, and  
\beq
x_{h} = \frac{E_h}{(E_{h} + E_{\nu h})}, \quad 
x_{l} = \frac{E_l}{(E_{l} + E_{\nu l})}. 
\eeq
Our event selection prescription involves four independent 
parameters which we vary in order to optimize the 
signal significances. These are 
the minimum transverse momentum $p_T$ of the hadronic 
$\tau$-lepton ($p^{\tau_{h}}_{T}$), minimum missing 
transverse energy ($\EMETmin$), minimum transverse 
momentum of the two leading jets ($P^{\rm min}_{T,j}$) 
and minimum of the di-jet invariant mass 
($m^{\rm min}_{j_{1}j_{2}}$). We proceed in the following steps.

\begin{itemize}
\item C1: We demand the presence of exactly one lepton (electron or muon) 
      with $p^{e}_{T} > 25$~GeV or $p^{\mu}_{T} > 20$~GeV.
\item C2: We identify hadronic $\tau$ with $p_{T} > p^{\tau_{h}}_{T}$ and charge 
      opposite to that of the identified lepton.
\item C3: We select events with missing transverse energy $\EMET$ greater 
than  $\EMETmin$.
\item C4: The variables associated to di-tau invariant mass reconstruction satisfy $0 \le x_{l} \le 0.75$, $0 \le x_{h} \le 1$, and $\cos{\Phi_{lh}} \ge$ -0.9. 
\item C5: A cut on the transverse mass($m_{T}$) of the lepton and $\EMET$ is applied 
      to suppress the $W + {\rm jets}$ and $t \bar t$ backgrounds, where 
      \begin{equation}
        m_T^2 = 2 p_T^{l} \EMET ( 1 - \cos{\Delta\Phi}), 
      \end{equation}
      with $p_{T}^{\rm lep}$ representing the transverse momentum of the lepton and 
     $\Delta\Phi$ is the angle between that lepton and $\EMET$ in the transverse plane. 
     We demand $m_{T} <$ 30 GeV.  
\item C6: We require the leading two jets to satisfy $p_{T} \ge$ $p^{\rm min}_{T,j}$.
\item C7: The forward jets should lie in opposite hemispheres $\eta_{j_1} \times \eta_{j_2} \le$ 0 with 
      tau centrality ${\rm min}({\eta_{j_1},\eta_{j_2}})\le \eta_{{\rm lep},{\tau}} \le 
{\rm max}({\eta_{j_1},\eta_{j_2}})$ for the two highest $p_T$ jets.
\item C8: Forward jets should also satisfy 
$\Delta\eta_{j_{1}j_{2}} \ge$ 4.4 and di-jet invariant mass 
      $m_{j_{1}j_{2}} \ge$ $m^{\rm min}_{j_{1}j_{2}}$.
\item C9: The events are rejected if there are any additional jets with $p_{T} \ge$ 20 GeV 
      with $|\eta| \le$ 3.2.
\item C10: Finally, we select the events with di-tau invariant mass satisfying 
      90 GeV $\le$ $m_{\tau\tau} \le$ 110 GeV.

\end{itemize}

\begin{table}[!htb]\centering
\begin{tabular}{|c|cccc|}
\hline
Signal Regions & $p^{\tau_{h}}_{T}$ &$\EMETmin$ & $p^{{\rm min}}_{T,j}$ & $m^{\rm min}_{j_{1}j_{2}}$\\
\hline
SR0 & 20.0 & 30.0 & 20.0  & 700.0 \\
SRA & 80.0 & 80.0 & 60.0  & 1200.0 \\
SRB & 50.0 & 50.0 & 40.0  & 1500.0 \\
\hline
\end{tabular}
\caption{ \small
{\it Details of the different signal regions are given. 
The kinematic selection cuts obtained from the 
ATLAS simulation is denoted by SR0, while our 
optimized selection cuts are described by SRA and SRB. For details 
see the text. 
}}
\label{tab:optcuts}
\end{table}

\noindent 

The possible SM backgrounds in this case come from
 $t\bar t+{\rm jets}$, $W + {\rm jets}$, $Z +{\rm jets}$ and  
di-boson final states ($WW$, $ZZ$, $WZ$). From the ATLAS simulation \cite{Aad:2009wy}, we find that 
$Z +{\rm jets}$ is the most dominant background with $Z \to \tau^{+}\tau^{-}$.
Hence, in our analysis, we simulate only the $Z +{\rm jets}$ background. 
We use 
MADGRAPH5 (v1 2.2.2) \cite{Alwall:2014hca} to generate the background events and then 
hadronize the events using PYTHIA (version 6.4.28) \cite{Sjostrand:2006za}. 
Table \ref{tab:optcuts} shows the optimized set of cuts.
The expected number of signal and background events at the 14 TeV
run of the LHC with 3000 ${\rm fb}^{-1}$ of integrated luminosity within a mass 
window of 90-110 GeV and the resulting statistical significances for three 
optimized scenarios are presented 
in Table~\ref{tab:signi1}. We obtain a maximum of 1.9$\sigma$ significance 
for the signal region SRA with 3000 ${\rm fb}^{-1}$ of data. 
Note that, we have considered only the 
dominant background $Z + {\rm jets}$. However, we expect the significances to be reduced even further
if we consider all the possible backgrounds. Thus, 
we conclude that the possibility of observing the 98 GeV Higgs boson 
via VBF process at the 14 TeV LHC run is rather poor
even at an integrated luminosity of $3000~{\rm fb}^{-1} $.

\begin{table}[!htb]\centering
\begin{tabular}{|c|ccc|}
\hline
& SR0 & SRA & SRB  \\
\hline
Signal (S) & 271.3 & 25.4   & 71.0  \\
\hline 
Backgounds (B) & 1713.3 & 55.3  &  221.1 \\
\hline 
Significance ($\mathcal S = {S \over \sqrt{B + \kappa^{2} B^{2}}}$) & 0.78 & 1.91 & 1.52  \\
\hline 
\end{tabular}
\caption{ \small
{\it
Expected number of events at the 14 TeV LHC with 
3000 ${\rm fb}^{-1}$ of integrated luminosity within a mass window
90 - 110 GeV for individual signal and total background. 
We assume that the signal cross section is 20\% of the 
SM value calculated with the Higgs mass $m^{\rm SM}_{h}$ = 98 GeV. 
The signal significances are calculated 
using systematic uncertainty $\kappa = 0.2$.}}
\label{tab:signi1}
\end{table}

\subsection{Direct search: 2. Associated production}

Another important production mode of the 98 GeV Higgs 
boson is the Higgs-strahlung process where the Higgs 
is produced in association with a gauge boson $W/Z$. In 
our earlier work \cite{older}, we discussed the discovery 
potential of the 98 GeV Higgs boson produced via this 
process giving special attention to the boosted regime 
with the assumption that it was produced with $p_T > 200 $ GeV. 
It is pertinent 
to note that in Ref.\cite{older} we took the number of 
background events directly from the ATLAS 
simulation \cite{ATLAS:2009elr}. However, in this work, we  
perform a more detailed Monte Carlo simulation by 
generating both the signal and background events and then 
optimizing the event selection cuts. We again focus on the 
boosted regime here. Note that, even though the production  
cross-section is very small in this highly boosted regime 
($p_T$ of the Higgs is greater than 200 GeV), 
relatively large kinematic acceptance and large background 
reduction make this analysis special. The details of our 
collider analysis can be found in Ref.~\cite{older}. However, 
we give a very brief outline of our analysis below. We divide this
part of our analysis 
into three categories based on the decay 
modes of the gauge bosons, namely, 

\begin{enumerate}
\item $Wh$ process with $W$ decaying leptonically with missing 
transverse momentum $\PMET > \PMETmin$ and 
$ p_T ^{e /\mu} >30$ ~{\rm GeV}. The transverse momentum 
of the $W$-boson must also satisfy $ p_T > p_{T,W}^{\rm min} $. 
Here we vary the quantities $\PMETmin$ and $p_{T,W}^{\rm min}$ 
independently.

\item $Zh$ process with $Z$ decaying into a pair of leptons (e/$\mu$) with 
di-lepton invariant mass satisfying $80~{\rm GeV} < m_{\ell \ell} < 100$ GeV 
while $ p^{Z}_T$ exceeds certain minimum value $p_{T,Z}^{\rm min}$. 
We also vary the transverse momentum of the two leptons independently.  

\item Finally, missing transverse energy driven signal with 
no leptons and $ \EMET > \EMETmin $. This kind of signature 
mainly comes from the process $Zh$ when $Z$-boson decays invisibly 
to a pair of neutrinos. However, contributions from $Wh$ process 
may also come when the lepton from $W$ remains undetected. 
\end{enumerate}

\noindent
In the above, $p_T^{\rm min}$ refers to the minimum $p_T$ 
of the Higgs required to claim the jet to be a {\it{Fat jet}} 
with $|\eta|<$~2.5. Similar to our VBF analysis, we first analyze 
the ATLAS optimized cuts and then vary some of the important 
observables relevant to a specific process so as to obtain 
the maximum sensitivity of the given channel. For example, 
for the $\ell^{+}\ell^{-}b\bar b$ channel we vary the transverse 
momentum of the Higgs jet, $Z$ boson and the pair of leptons 
to get the maximum signal significance. Similar strategy has 
been opted for other modes. The default selection cuts 
for a given channel have been denoted as ``SR0", while our 
optimized selection cuts are represented as ``SRA", ``SRB" 
and ``SRC" for the $\ell^{+}\ell^{-}b\bar b$, 
$\ell^{+}\nu_{\ell}b\bar b$ and $\EMET b\bar b$ signals respectively 
(see Table~\ref{tab:cuts}). We use 
PYTHIA (version 6.4.28)\cite{Sjostrand:2006za} for generation 
of signal events while FASTJET (version 3.0.3)\cite{Cacciari:2011ma} is used for 
reconstruction of jets and also implementation of the jet 
substructure analysis. The most dominant SM backgrounds for our 
process of interest are $WW$, $ZZ$, $WZ$, $Wb\bar b$, $Zb\bar b$ and 
$t\bar t$. We use MADGRAPH to generate the $Wb\bar b$, $Zb\bar b$ samples 
and then passed them to PYTHIA for showering and hadronization while the rest 
of the background samples are generated using PYTHIA itself.

We present our final results in Table~\ref{tab:sig1} where 
we scale the cross-sections by 0.2 and focus in the mass 
window 90 - 110 GeV to satisfy the 2.3$\sigma$ LEP excess.
In Table~\ref{tab:sig1} we display 
the expected number of events
for the signal and various backgrounds
with 3000 ${\rm fb}^{-1}$ of luminosity. The statistical 
significances for the signal regions SR0 (the default 
ATLAS parameters) as well as SRA, SRB and SRC (our optimized sets) 
are also shown for the three possible decay modes. We find 
that the best sensitivity comes from the channel 
$\ell^{+}\ell^{-}b\bar b$ with 2.6$\sigma$ significance, while 
$\ell^{+}\nu_{\ell}b\bar b$ and $\EMET b\bar b$ have 
statistical significances of 2.5$\sigma$ and 1.5$\sigma$ 
respectively. We must note that although we assume 20\% 
systematic uncertainty (i.e., $\kappa$ = 0.2), 
nevertheless with 3000 ${\rm fb}^{-1}$ of luminosity we can expect to 
have a better control over the various sources of 
systematic uncertainties leading to an enhanced 
signal significance. Therefore, one can expect to marginally 
exclude the 98-125 GeV Higgs scenario using 3000 ${\rm fb}^{-1}$ of 
data at the 14 TeV LHC via the Higgstrhlung process. 

\begin{table}[!htb]\centering
\begin{tabular}{|c |c|c cc cc|}
\hline
Process & Signal Regions & \multicolumn{5}{c|}{Selection Cuts} \\
\cline{1-7}
                &       & $p^{\rm min}_{T,Z}$    & $p^{\rm min}_{T,j_1}$   &  $\Delta\Phi_{1}$  & $p^{\ell_1}_{T}$   & $p^{\ell_2}_{T}$  \\
\cline{3-7}
 $\ell^{+}\ell^{-}b\bar b$       & SR0  &  180   &   200  & 1.2  & 25  & 20             \\
                 & SRA    & 250  & 250   & 1.2    & 100   & 50         \\
\hline
                &       &  $p^{\rm min}_{T,W}$    & $p^{\rm min}_{T,j_1}$   &  $\Delta\Phi_{1}$  & $\PMETmin$ &   \\
\cline{3-7}
$\ell^{+}\nu_{\ell}b\bar b$    & SR0  & 180  & 200 & 1.2 & 30       &              \\
                 & SRB        & 300      & 300        & 1.5   & 75       &          \\
\hline
                &       &   $p^{\rm min}_{T,j_1}$   &  $\Delta\Phi_{1}$  & $\PMETmin$  & &  \\
\cline{3-7}
$\EMET b\bar b$     & SR0        & 200  & 1.2   & 200   &        &              \\
                 & SRC        &  250  & 1.5  & 300   &          &          \\
\hline
\end{tabular}
\caption{ \small
{\it Different signal regions for the three processes 
considered in this analysis. The default ATLAS simulation 
is denoted by SR0, while our
optimized selection cuts are described by SRA, SRB and SRC for the 
three processes. }
}
\label{tab:cuts}
\end{table}

\begin{table}[!htb]\centering
\begin{tabular}{|c|cccc|}
\hline
Process & Signal Region & Signal (S) & Background (B) & Significance ($ S \over \sqrt{B + \kappa^{2} B^{2}}$)
\\
& & & & \\
\hline
$ \ell^+ \ell^- b \bar{b}$ & SR0 & 184.9  & 956.1 & 0.95  \\
& SRA & 94.8 & 168.7 & 2.62   \\
\hline
$ \ell \nu b \bar{b}$ & SR0 & 360.2  & 1998.4 &  0.89  \\
& SRB & 99.5 & 185.8 & 2.51 \\
\hline
$ \EMET b \bar{b}$ & SR0 & 184.9  & 2614.3 & 0.35  \\
& SRC & 94.8 & 297.2  & 1.53   \\
\hline
\end{tabular}
\caption{ \small
{\it
Expected number of events at the 14 TeV LHC run 
with 3000 ${\rm fb}^{-1}$ of integrated luminosity within a mass window
of 90 - 110 GeV for the individual signal and 
the combined background processes assuming 20\% LEP excess in this region of interest.}
}
\label{tab:sig1}
\end{table}


\subsection{Indirect search: Higgs coupling measurements }

\begin{figure}[!htb]
 \begin{center}
 {\includegraphics[angle =0, width=0.42\textwidth]{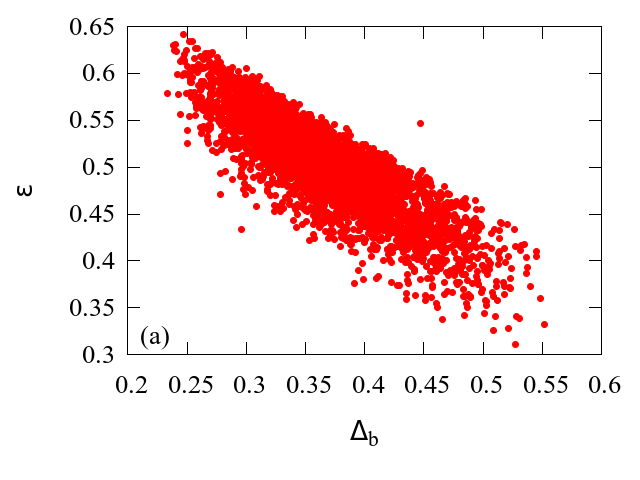} }
  {\includegraphics[angle =0, width=0.42\textwidth]{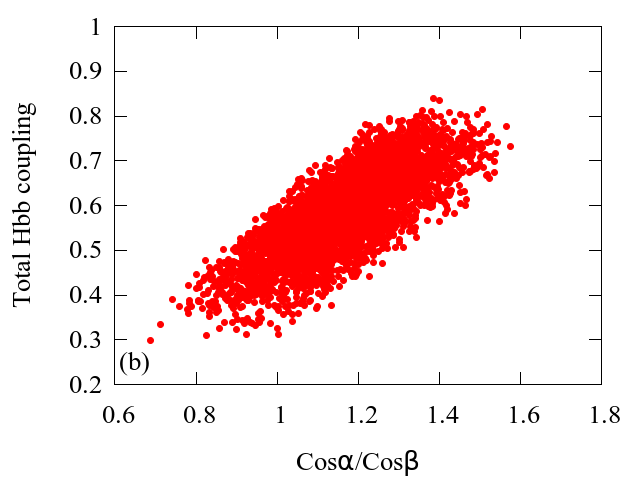} }
\caption{{\small Left panel shows the distribution of the quantity 
$\epsilon = \biggl({1\over 1+\Delta_b}\biggr) \times \biggl(1+{\Delta_b \cot\beta \tan \alpha}\biggr)$
with the variation of $\Delta_b$. The variation of the total 
Yukawa coupling of the bottom quark with the heavier Higgs boson
with respect to the tree level 
coupling $\frac{\cos\alpha}{\cos\beta}$ is shown in the right panel.}}  
\label{fig:c1_c2c3}
 \end{center}
\end{figure}

Precise measurement of the various Higgs signal 
strength variables can prove to be significant to 
indirectly probe/exclude the ILLH scenario. 
We find that the important 
observables which can play crucial roles in this regard are 
${\mu}_{bb}$, ${\mu}_{\tau \tau}$ and ${\mu}_{ZZ}$ for a 
given production mechanism of the observed 
125 GeV Higgs boson. Before going into a detailed analysis of these observables, 
let us first discuss some of the important Higgs boson 
couplings. The tree level Yukawa 
coupling of the bottom quark with the heavy Higgs boson of the MSSM having a mass 
around 125 GeV goes 
as $\cos\alpha \over \cos\beta$ where $\tan\beta$ is the ratio of 
the vevs of the two Higgs 
doublets. However, loop corrections (in orders 
of $\alpha_s\tan\beta$) involving various supersymmetric particles can 
significantly modify the tree level $Hb\bar b$ Yukawa coupling. These 
effects are generally denoted by the quantity $\Delta_b$ and 
the additional contribution coming from $\Delta_b$ can be 
summarized as \cite{Carena:1999py,Hall:1993gn,Guasch:2003cv,Dawson:2011pe},  
\beq
\epsilon = \biggl({1\over 1+\Delta_b}\biggr) \times \biggl(1+{\Delta_b \cot\beta \tan \alpha}\biggr).  
\eeq
In the left panel of Fig.~\ref{fig:c1_c2c3}, we show the dependence on
$\Delta_b$ of the quantity $\epsilon$ which estimates the loop 
contribution to the tree level Yukawa coupling. We find that the effect of 
$\Delta_b$ is indeed significant in the parameter space of our interest. 
In the right panel of Fig.~\ref{fig:c1_c2c3}, we show the 
variation of the complete $Hb\bar b$ coupling (the effect of 
$\Delta_b$ included) with the 
tree level coupling. It is evident 
from both the figures that for a significant number 
of points $\Delta_b$ is indeed large. Thus, even though 
the maximum value of the tree level coupling goes up to 
1.4-1.5, the total $Hb\bar b$ coupling never exceeds unity, 
implying that the $Hb\bar b$ coupling is always suppressed with respect to 
the SM expectation which can indeed serve as a very 
distinctive feature of the ILLH scenario.

\begin{figure}[!htb]
 \begin{center}
 {\includegraphics[angle =0, width=0.42\textwidth]{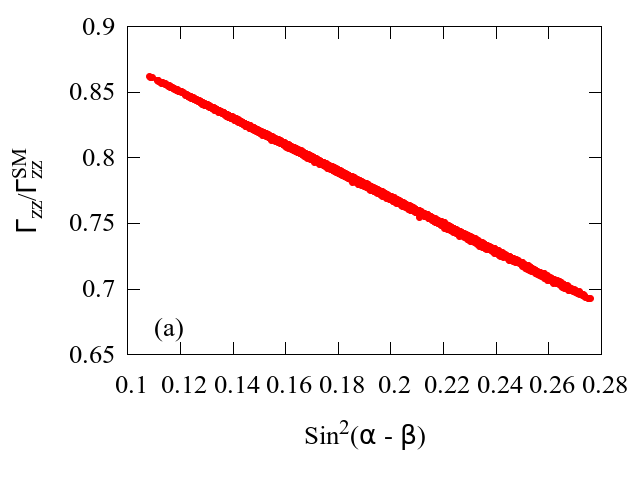} }
 {\includegraphics[angle =0, width=0.42\textwidth]{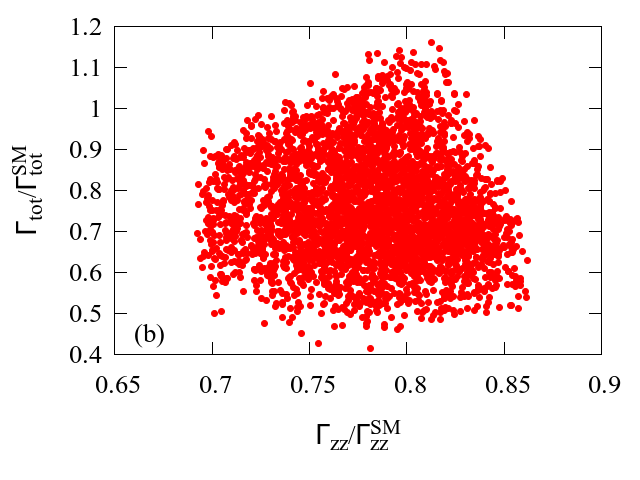} }
\caption{{\small The correlation of the partial decay width of Higgs to 
$ZZ^*$  decay with $\sin^{2}({\beta - \alpha})$ (Left panel) 
and with total Higgs decay width $\Gamma_{tot}$ (Right panel).}}
\label{fig:sin2alpha_hzzratio}
 \end{center}
\end{figure}

Let us now turn our attention to another important 
decay mode $H \to ZZ^{*}$. In the left panel of 
Fig.~\ref{fig:sin2alpha_hzzratio}, we show 
the variation of the ratio $\Gamma_{ZZ}/\Gamma_{ZZ}^{SM}$ 
to the square of the tree level $HZZ$ coupling $\sin({\beta - \alpha})$.
Here, $\Gamma_{ZZ}$ denotes the partial width of the decay $H \to ZZ^{*}$ 
in the MSSM and $\Gamma_{ZZ}^{SM}$ denotes the same for the SM. The 
behaviour is well understood; as the coupling decreases so does the partial width.
However, we must note that 
the $H \to ZZ^*$ partial width 
is also suppressed in this case. With both the partial widths for 
the decays $H \to b\bar b$ and $H \to ZZ^*$ 
suppressed, one can expect to observe a significant suppression in 
the total decay width of the Higgs boson as well. From the plot in the right panel of  
Fig.~\ref{fig:sin2alpha_hzzratio}, where the X-axis denotes the ratio 
$\Gamma_{ZZ}/\Gamma_{ZZ}^{SM}$ and the Y-axis stands for 
the ratio of the total Higgs decay width ($\Gamma_{tot}$) in the 
MSSM to that in the SM ($\Gamma_{tot}^{SM}$), one can easily observe 
the suppression in the total decay width. Thus, one may also expect 
to find a mild enhancement in partial widths of the sub-dominant decay modes like 
$\tau^{+}\tau^{-}$, $gg$.

\begin{figure}[!htb]
 \begin{center}
{\includegraphics[angle =0, width=0.32\textwidth]{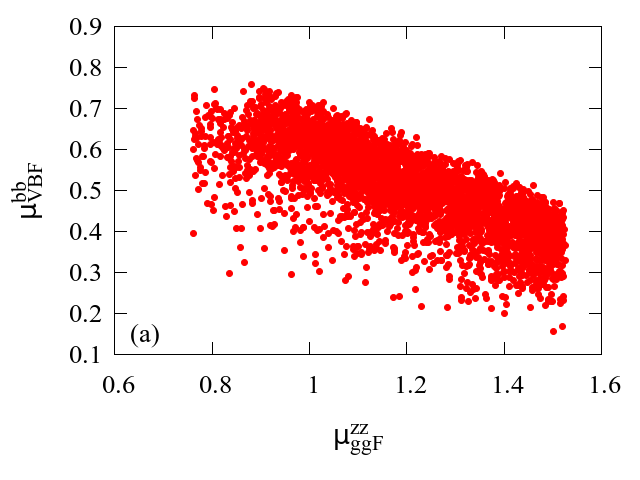} }
{\includegraphics[angle =0, width=0.32\textwidth]{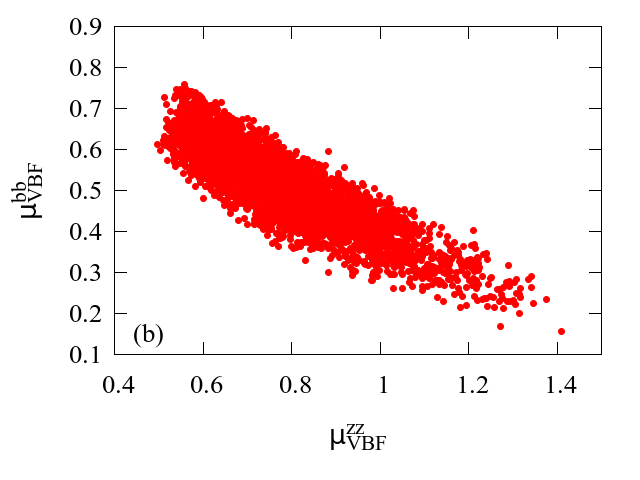} }
{\includegraphics[angle =0, width=0.32\textwidth]{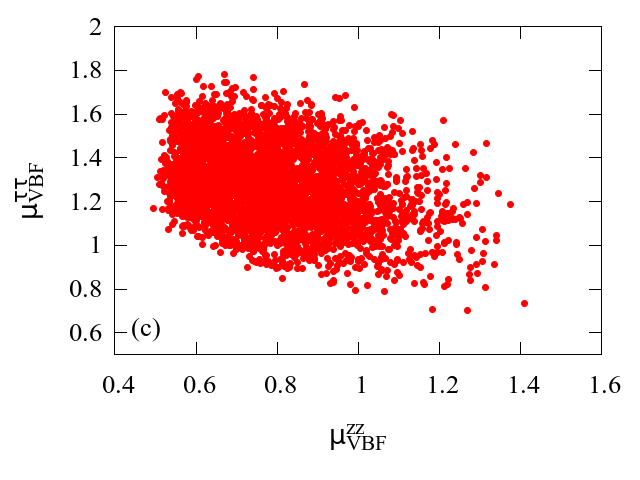} }
\caption{{\small Correlation plots in ${\mu}^{ZZ}_{ggF} - {\mu}^{bb}_{VBF}$ 
(left), ${\mu}^{ZZ}_{VBF} - {\mu}^{bb}_{VBF}$ (middle) and 
${\mu}^{ZZ}_{VBF} - {\mu}^{\tau\tau}_{VBF}$ (right) planes. 
For details see the text.}}
\label{fig:corrln}
 \end{center}
\end{figure}

\begin{figure}[!htb]
 \begin{center}
{\includegraphics[angle =0, width=0.42\textwidth]{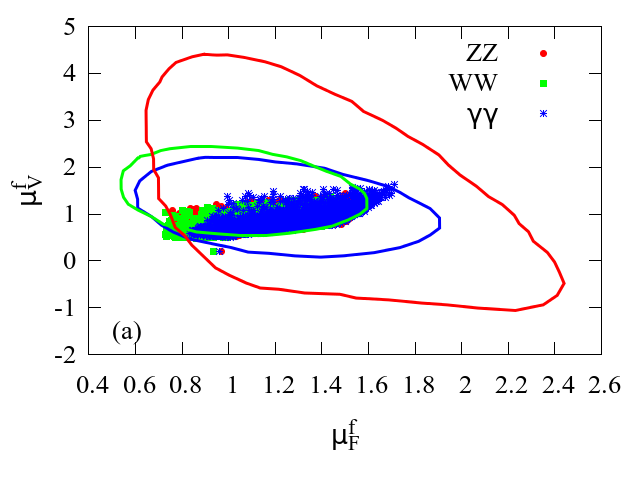} }
{\includegraphics[angle =0, width=0.42\textwidth]{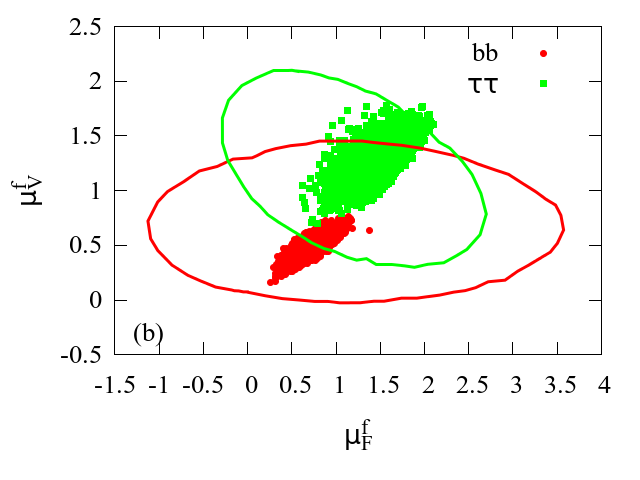} }
\caption{{\small The 95\% C.L. contours in the 
$\mu^{f}_{ggF+t\bar t H}$ - $\mu^{f}_{VBF+VH}$ plane with the 
ATLAS and CMS combined 7 and 8 TeV data for five possible decay 
modes: in the left we show $\gamma\gamma$, $ZZ$ and 
$WW$ channels while right panel for $b\bar b$ and $\tau^{+}\tau^{-}$. 
The subscript `F' denotes the combined ggF and $t\bar t H$ process, while
`V' signifies the combined VBF and VH processes. However, here for the 
``fusion" (F) mode we consider ggF only as Higgs production 
via ggF is much larger compared to the $t\bar t H$ process. See the text for 
more details.}}
\label{fig:muV_muF}
 \end{center}
\end{figure}

Improved measurement of the signal strength variables 
at the high luminosity run of the LHC may help us to probe 
the ILLH scenario indirectly. We present a detailed study in this regard
as follows.
We have already introduced the signal strength variables 
in Sec.~\ref{sec2}. Here, we discuss the variables which we 
find interesting and which are seen to have some impact to probe/exclude the 
parameter space of our interest. In the left-most panel of 
Fig.~\ref{fig:corrln}, we display the correlation 
in the ${\mu}^{ZZ}_{ggF} - {\mu}^{bb}_{VBF}$ 
plane. We assume gluon-gluon fusion (ggF) process as the production 
mechanism for the Higgs boson decaying into $ZZ$ while for the final state 
$b\bar b$ the Higgs is taken to be produced via vector-boson fusion (VBF) 
process. We find that although ${\mu}^{ZZ}_{ggF}$ can vary in the range
0.6 - 1.6, the values of ${\mu}^{bb}_{VBF}$ is seen to be
suppressed i.e., less than unity. Furthermore, even 
though gluon-gluon fusion process is the dominant one, 
associated production mechanism of the Higgs can also be used to measure 
these signal strength variables. We discuss the correlations of 
three such variables, namely ${\mu}^{ZZ}_{VBF}$, ${\mu}^{bb}_{VBF}$ 
and ${\mu}^{\tau\tau}_{VBF}$. 
In the middle and right-most panel we show the correlations 
in the ${\mu}^{ZZ}_{VBF} - {\mu}^{bb}_{VBF}$ and 
${\mu}^{ZZ}_{VBF} - {\mu}^{\tau\tau}_{VBF}$ planes 
respectively. Similar to the ggF case, there are no such restrictions
on the values of ${\mu}^{ZZ}_{VBF}$. However, we find a strong anti-correlation 
between ${\mu}^{bb}_{VBF}$ and ${\mu}^{\tau\tau}_{VBF}$. 
The values of ${\mu}^{bb}_{VBF}$ is found to be always less than 
unity while those of ${\mu}^{\tau\tau}_{VBF}$ to 
be dominantly greater than unity. {At this point, one 
might be interested to know the present status of the measurement 
of these Higgs signal strength correlations at the LHC. In 
Fig.~\ref{fig:muV_muF}, we study these correlations in 
the $\mu^{f}_{ggF+t\bar t H}$ - $\mu^{f}_{VBF+VH}$ plane for a 
generic final state $f$, and then 
compare the model predictions with the 95\% correlation contours 
obtained using the 10-parameter fit for the five decay modes 
of the observed Higgs boson by the ATLAS and CMS combined 
7 and 8 TeV data \cite{ATLAS-CMS-comb}. In the left panel, 
we show $\gamma\gamma$, $ZZ$ and
$WW$ channels while right panel for $b\bar b$ and $\tau^{+}\tau^{-}$. 
The subscript `F' denotes the combined ggF and $t\bar t H$ process, while
`V' signifies the combined VBF and VH processes. However, here for the 
``fusion" (F) mode we consider ggF only as Higgs production 
via ggF is much larger compared to the $t\bar t H$ process. 
Comparing the correlation plots (Fig.~\ref{fig:muV_muF}) 
with Table \ref{tab:higgs}, we find that 
at present the impact of these correlations at the parameter 
space of interest is comparable with 
that of the individual signal strengths (less than 2\% points 
are found to be outside the 95\% C.L. contours). However, 
here we would like to note that with precise measurements in the 
future runs of LHC these contours are expected to shrink, this may 
lead to interesting consequences for our model parameter space.} 
Considering the future improvements in signal strength measurements, 
if we assume that at the high luminosity run of the LHC with 
3000 ${\rm fb}^{-1}$ of data ${\mu}^{ZZ}_{VBF}$ can be measured 
with an accuracy at the level of 30\%, then from these 
correlations we can infer that ${\mu}^{\tau\tau}_{VBF}$ will always 
have values larger than unity. However, ${\mu}^{bb}_{VBF}$ will be 
less than 0.8. Thus, for a given measurement 
of the signal strength variable in the $ZZ$ channel, observation 
of suppression of the same quantity for the $b\bar b$ channel and 
enhancement in the 
$\tau\tau$ channel will be an ideal probe of the ILLH scenario.

Before we end this section, let us summarize our results from 
the collider analysis. We analyze the most important production 
mechanism of the 98 GeV Higgs boson, namely the VBF and associated 
production processes, and find that these processes are not 
sensitive enough to exclude the ILLH scenario. We then attempt to 
exclude this possibility indirectly using various Higgs signal 
strength variables, and here we find very distinctive features in the 
correlations of the signal strength variables for $bb$, $\tau\tau$ 
and $ZZ$ decay channels. However, there exists several  
other processes which can also be used directly to probe this scenario,
e.g. the associated production of the 98 GeV Higgs boson 
with a pair of top quarks ($t\bar t h$). Even though this process 
do not directly involve the $\sin^2({\beta - \alpha})$ coupling, still 
it can be used to probe this scenario. In our earlier work \cite{older},  
we performed a naive collider analysis of this process, following the 
same analysis for a SM Higgs boson \cite{Plehn:2009rk}, and obtained 
a 2.6$\sigma$ statistical significance without considering 
systematic uncertainties. A more detailed study is required, specially 
focusing on the boosted regime and applying the jet substructure technique, 
which is beyond the scope of the present work. Interestingly, we can also 
discover/exclude the 98-125 GeV Higgs scenario by looking for the other 
Higgses present in this model, e.g. the neutral 
CP-odd Higgs $A$ and the charged Higgs bosons $H^{\pm}$. One 
of the most important characteristic signatures of this ILLH scenario 
is the presence of relatively light $A$ and $H^{\pm}$ bosons with masses 
$\lsim$ 200 GeV. The pseudoscalar Higgs $A$ is produced via 
gluon-gluon fusion  and/or $b\bar b$ fusion process and, 
if kinematically allowed, can decay to $Zh$ giving rise interesting 
final state topologies involving multi-leptons and multiple b-jets. 
On the other hand, the charged Higgs bosons are produced 
by $tbH^{\pm}$ process and 
decays dominantly into $\tau\nu_{\tau}$ and $t\bar b$ final 
states. A dedicated analysis 
for these very light CP-odd and charged Higgses in the context 
of high luminosity run of the LHC is required. We leave this very 
interesting possibility for our future work. Thus, {\it the observation of 
a charged Higgs and a pseudoscalar Higgs boson with masses 
between 140 - 200 GeV and simultaneously the non-observation of 
any CP-even Higgs in the same region will be 
a direct probe of the ILLH scenario}. Furthermore, productions of 
these light Higgses in pair, e.g. processes 
like $H^{\pm}A$, $H^{+}H^{-}$, $H^{\pm}h$, $Ah$ are also very 
interesting possibilities at the LHC. There exists a study considering 
all of these processes at the LHC in the context of non-decoupling 
region of the MSSM \cite{Christensen:2012si}. However, the authors 
focused only on the regions with 
masses of these Higgses lying between 95 - 130 GeV. Note that, 
this 95 - 130 GeV region 
is already excluded by the pseudoscalar and charged Higgs searches at the LHC, and 
thus a study focussing on the region 140 - 200 GeV is now required which we plan to 
address in our future correspondence.  We would like to mention that in our earlier work
we discussed the possibility of observing the 98 GeV Higgs boson 
directly at the ILC. We found that the above could be easily 
discovered/excluded at the 250 GeV ILC with a 100 ${\rm fb}^{-1}$ of luminosity
which is easily achievable within the first few years of
its run. {{ Finally we would like to add one important point regarding 
the direct measurement of various SUSY particles at the high luminosity (HL)
run of LHC. The expected exclusion limits at the HL-LHC 
for the top and bottom squarks are around 1 TeV, for charginos and neutralinos 
around 600 GeV while for gluinos around 2.5 TeV \cite{Cakir:2014nba}. We 
check that even with such a heavy sparticle spectrum, there is ample parameter 
space which satisfies the current LHC data, thus we conclude that it 
is almost impossible to exclude the 98-125 GeV scenario even at the 
high luminosity run of LHC.}}


\section{Summary and conclusions}
\label{sec5}

The objective of this work is to interpret the LEP excess observed around
a mass of $\sim$ 98 GeV as the lighter CP-even Higgs boson of the MSSM while
the LHC-observed scalar at $\sim$ 125 GeV plays the role of the heavier
one. We analyze this scenario in the light of the latest results from 
the LHC including the limits on Higgs signal strengths. Other 
relevant constraints like those coming from the flavor 
sector e.g. $\bsg$, $\bsmumu$ and the DM relic density constraint
are also taken into account.
The ATLAS/CMS searches in the $H/A \rightarrow \tau^+ \tau^-$ and 
charged Higgs searches restrict the values of 
$\tan\beta$ to a very narrow range. By performing a detailed random 
scan over the MSSM parameter space we try to pin point the region
of parameter space where all the above constraints can be 
simultaneously accommodated. We observe that the ILLH scenario
can still be harboured within the MSSM framework. The values of $\mu$
required to satisfy the above criteria are generically large $\gsim$ 
7 TeV and $A_t$ tends to assume only appreciably large positive values.

To perform the analysis in a model independent way we use the limits
on $\sigma \times {\rm BR}(\Phi \rightarrow \tau^+ \tau^-)$, $\sigma$ being
the production cross-section of the non-standard Higgs boson $\Phi$.
The LHC limits are available for ggF and VBF production processes.
We observe that  $\sigma (ggF) \times {\rm BR}(\Phi \rightarrow \tau^+ \tau^-)$
lies well below the experimental limit for the small values of $\tan\beta$
considered in this analysis. However, the values of this observable for 
associated production with $b \bar b$ seems to be very close to the present 
experimental bound.

In the ILLH scenario that is associated with a light $H^\pm$, the constraint 
from $\bsg$ plays an important role to eliminate a large region 
of parameter space. We remember that the associated  
SM contributions almost saturate the the experimental limit. 
Generally, the $\chonepm-\tilde t_1$  loop contributions are not 
large enough to effectively cancel the contributions from the $H^\pm-t$ 
loops where $H^\pm$ is light. 
This causes a large amount of the parameter space to be discarded. 
It is only for large $\mu$ zone (with large $A_t$ values) along with 
sign$(\mu A_t)>0$
the NLO contributions arising out of the top-quark Yukawa coupling 
partially cancel the leading order contributions of the $H^\pm-t$ loops. 
Additionally, there is an overall suppression coming out of SQCD 
corrections to the mass $m_b$.  Thus the available parameter space 
that satisfies $\bsg$ constraint has 
large values of $\mu$ with sign$(\mu A_t)>0$. 
The other constraint namely $\bsmumu$ is not a very 
stringent one in this zone of parameter space that survive the $\bsg$ 
constraint.

An important result regarding the Higgs signal strength variables
is obtained when we closely study the points satisfying all the constraints along with the limits on $R^H_{gg}(\gamma \gamma)$.
If we further demand that values of $R^H_{gg}(ZZ)$ lie within 20\% around
the SM value of unity, all the points are seen to have $R^H_{VH}(bb) \lsim 0.8$.
As already mentioned, the loop correction to bottom quark 
Yukawa coupling and hence to bottom quark mass ($\Delta m_b$) is significantly
large in the present case. This reduces the partial decay width
$\Gamma_{(H \rightarrow b \bar b)}$. The value of 
$\Gamma_{(H \rightarrow ZZ)}$ is also small, leading to a reduction
in the total decay width. However, the ${\rm BR}(H \rightarrow ZZ)$
can be significantly large. Thus, for the points with 
$0.8 < R^H_{gg}(ZZ) < 1.2$ the value of $R^H_{VH}(bb)$ is 
seen to be $\lsim 0.8$. This can play a major role as a distinctive 
feature of the present scenario 
provided the sensitivity on the coupling strength measurements is increased
to a desired accuracy. 

We analyze the possibilities of observing the ILLH scenario in the 14 TeV run
of the LHC in two production channels, the vector boson fusion process
and associated production with $W/Z$ boson. For the VBF process we follow
the ATLAS simulation for the decay mode $H \rightarrow \tau_l \tau_h$.
The selection cuts are varied to obtain an optimum signal to background
ratio. We simulate only the $Z$ + jets events as the dominant SM background
in this analysis.  From statistical significances of the three optimized 
scenarios considered here we conclude that the possibility of observing
the ILLH scenario with 3000 ${\rm fb}^{-1}$ of luminosity is rather small.

For the Higgstrahlung processes we concentrate on the highly boosted
regime where the Higgs boson of mass 98 GeV is produced with 
 $p_T > $ 200 GeV. Three different scenarios are
considered here depending on the decays of the associated gauge boson. 
These are the $W H$ process with $W$ decaying leptonically,
the $Z H$ process with $Z$ decaying to $e/\mu$ pair, the $ZH$
process with invisible decays of the $Z$ boson.
We generate both the signal and background events 
through a detailed Monte Carlo simulation following the 
ATLAS analysis. 
The results are presented for three optimized selection regions
along with the one using default selection cuts. From the
results we observe that the ILLH scenario can be marginally
ruled out with 3000 ${\rm fb}^{-1}$ of luminosity at the 14 TeV run of the LHC.

Finally, we summarize our findings of the 98 - 125 GeV ILLH 
scenario as follows: 

\begin{itemize}

\item The most updated LHC data along with the low energy physics flavor data and bounds from dark matter searches 
    does not exclude the possibility of having a 98 - 125 GeV Higgs scenario. We provide two sample benchmark 
    points in support of our results.  
     
\item The possibility for direct detection of the 98 GeV Higgs boson at the run-2 of the LHC is marginal even after 
    using the state-of-the-art jet substrcuture technique.

\item However precise measurements of the Higgs signal strengths may act as an indirect probe of the ILLH scenario.  
    We find interesting correlations between these signal strength variables. For example, the quantity $R_{b\bar b}$ 
    is always less that unity, thus we find that if we can measure the Higgs signal strength associated to the 
    Higgs decay to $ZZ$ then we must see a strong suppression in the $b\bar b$ mode. We expect at the 
    high luminosity run of LHC these measurements will be improved by few orders of magnitude, and thus could easily 
    be used as a probe of the ILLH scenario. 
\end{itemize}

\vskip 0.5cm
\noindent
{\large \bf \underline {Acknowledgments:}}
\vskip 0.12cm
\noindent
The work of BB is supported by Department of Science and 
Technology, Government
of INDIA under the Grant Agreement numbers IFA13-PH-75 
(INSPIRE Faculty Award). 
AC would like to thank the Department of Atomic Energy,
Government of India for financial support. 
MC would like to thank Council of Scientific and Industrial Research,
Government of India for financial support. We gratefully acknowledge the help 
received from Abhishek Dey for running the code Vevacious for the benchmark 
points.


\end{document}